# A Dust-Obscured Massive Maximum-Starburst Galaxy at a Redshift of 6.34

*Dominik A. Riechers[1,2], C.M. Bradford[1,3], D.L. Clements[4], C.D. Dowell[1,3], I. Pérez-Fournon[5,6], R.J. Ivison[7,8], C. Bridge[1], A. Conley[9], Hai Fu[10], J.D. Vieira[1], J. Wardlow[10], J. Calanog[10], A. Cooray[10,1], P. Hurley[11], R. Neri[12], J. Kamenetzky[13], J.E. Aguirre[14], B. Altieri[15], V. Arumugam[8], D.J. Benford[16], M. Béthermin[17,18], J. Bock[1,3], D. Burgarella[19], A. Cabrera-Lavers[5,6,20], S.C. Chapman[21], P. Cox[12], J.S. Dunlop[8], L. Earle[9], D. Farrah[22], P. Ferrero[5,6], A. Franceschini[23], R. Gavazzi[24], J. Glenn[13,9], E.A. Gonzalez Solares[21], M.A. Gurwell[25], M. Halpern[26], E. Hatziminaoglou[27], A. Hyde[4], E. Ibar[7], A. Kovács[1,28], M. Krips[12], R.E. Lupu[14], P.R. Maloney[9], P. Martinez-Navajas[5,6], H. Matsuhara[29], E.J. Murphy[1,30], B.J. Naylor[3], H.T. Nguyen[3,1], S.J. Oliver[11], A. Omont[24], M.J. Page[31], G. Petitpas[25], N. Rangwala[13], I.G. Roseboom[11,8], D. Scott[26], A.J. Smith[11], J.G. Staguhn[16,32], A. Streblyanska[5,6], A.P. Thomson[8], I. Valtchanov[15], M. Viero[1], L. Wang[11], M. Zemcov[1,3], J. Zmuidzinas[1,3]*

[1]California Institute of Technology, 1200 East California Blvd., MC 249-17, Pasadena, CA 91125, USA
[2]Cornell University, 220 Space Sciences Building, Ithaca, NY 14853, USA
[3]Jet Propulsion Laboratory, 4800 Oak Grove Drive, Pasadena, CA 91109, USA
[4]Astrophysics Group, Imperial College London, Blackett Laboratory, Prince Consort Road, London SW7 2AZ, UK
[5]Instituto de Astrofisica de Canarias, E-38200 La Laguna, Tenerife, Spain
[6]Departamento de Astrofisica, Universidad de La Laguna, E-38205 La Laguna, Tenerife, Spain
[7]UK Astronomy Technology Centre, Royal Observatory, Blackford Hill, Edinburgh EH9 3HJ, UK
[8]Institute for Astronomy, University of Edinburgh, Royal Observatory, Blackford Hill, Edinburgh EH9 3HJ, UK
[9]Center for Astrophysics and Space Astronomy 389-UCB, University of Colorado, Boulder, CO 80309, USA
[10]Dept. of Physics & Astronomy, University of California, Irvine, CA 92697, USA
[11]Astronomy Centre, Dept. of Physics & Astronomy, University of Sussex, Brighton BN1 9QH, UK
[12]Institut de RadioAstronomie Millimetrique, 300 Rue de la Piscine, Domaine Universitaire, F-38406 Saint Martin d'Heres, France
[13]Dept. of Astrophysical and Planetary Sciences, CASA 389-UCB, University of Colorado, Boulder, CO 80309, USA
[14]Department of Physics and Astronomy, University of Pennsylvania, Philadelphia, PA 19104, USA
[15]Herschel Science Centre, European Space Astronomy Centre, Villanueva de la Cañada, 28691 Madrid, Spain
[16]Observational Cosmology Lab, Code 665, NASA Goddard Space Flight Center, Greenbelt, MD 20771, USA
[17]Laboratoire AIM-Paris-Saclay, CEA/DSM/Irfu - CNRS - Universite Paris Diderot, CEA-Saclay, point courrier 131, F-91191 Gif-sur-Yvette, France
[18]Institut d'Astrophysique Spatiale (IAS), batiment 121, Universite Paris-Sud 11 and CNRS, UMR 8617, F-91405 Orsay, France
[19]Aix-Marseille Universite, CNRS, Laboratoire d'Astrophysique de Marseille, UMR7326, F-13388 Marseille, France
[20]Grantecan S.A., Centro de Astrofisica de La Palma, Cuesta de San Jose, E-38712 Brena Baja, La Palma, Spain
[21]Institute of Astronomy, University of Cambridge, Madingley Road, Cambridge CB3 0HA, UK
[22]Department of Physics, Virginia Tech, Blacksburg, VA 24061, USA
[23]Dipartimento di Fisica e Astronomia, Universita di Padova, vicolo Osservatorio, 3, I-35122 Padova, Italy
[24]Institut d'Astrophysique de Paris, UMR 7095, CNRS, UPMC Univ. Paris 06, 98bis boulevard Arago, F-75014 Paris, France
[25]Harvard-Smithsonian Center for Astrophysics, 60 Garden Street, Cambridge, MA 02138, USA
[26]Department of Physics & Astronomy, University of British Columbia, 6224 Agricultural Road, Vancouver, BC V6T 1Z1, Canada
[27]ESO, Karl-Schwarzschild-Str. 2, D-85748 Garching bei München, Germany
[28]Institute for Astrophysics, University of Minnesota, 116 Church Street SE, Minneapolis, MN 55455, USA
[29]Institute for Space and Astronautical Science, Japan Aerospace and Exploration Agency, Sagamihara, Kanagawa 229-8510, Japan
[30]Infrared Processing and Analysis Center, MS 100-22, California Institute of Technology, Pasadena, CA 91125, USA
[31]Mullard Space Science Laboratory, University College London, Holmbury St. Mary, Dorking, Surrey RH5 6NT, UK
[32]Department of Physics & Astronomy, Johns Hopkins University, Baltimore, MD, 21218, USA

**Massive present-day early-type (elliptical and lenticular) galaxies probably gained the bulk of their stellar mass and heavy elements through intense, dust-enshrouded starbursts – that is, increased rates of star formation – in the most massive dark matter halos at early epochs. However, it remains unknown how soon after the Big Bang such massive starburst progenitors exist. The measured redshift (*z*) distribution of dusty, massive starbursts has long been suspected to be biased low in redshift owing to selection effects,[1] as confirmed by recent findings of systems out to redshift *z*~5.[2,3,4] Here we report the identification of a massive starburst galaxy at redshift 6.34 through a submillimeter color-selection technique. We unambiguously determined the redshift from a suite of molecular and atomic fine structure cooling lines. These measurements reveal a hundred billion solar masses of highly excited, chemically evolved interstellar medium (ISM) in this galaxy, which constitutes at least 40% of the baryonic mass. A "maximum starburst" converts the gas into stars at a rate more than 2,000 times that of the Milky Way, a rate among the highest observed at any epoch. Despite the overall downturn of cosmic star formation towards the highest redshifts,[5] it seems that environments mature enough to form the most massive, intense starbursts existed at least as early as 880 million years after the Big Bang.**

We have searched 21 deg$^2$ of the *Herschel*/SPIRE data of the HerMES blank field survey[6] at 250 – 500 μm for "ultra-red" sources with flux densities $S_{250\mu m} < S_{350\mu m} < S_{500\mu m}$ and $S_{500\mu m}/S_{350\mu m}>1.3$, i.e., galaxies that are significantly redder (and thus, potentially at higher redshift) than massive starbursts discovered thus far. This selection yields five candidate ultra-red sources down to a flux limit of 30 mJy at 500 μm (>5$\sigma$ and above the confusion noise; see Supplementary Information Section 1 for additional details), corresponding to a source density of ≤0.24 deg$^{-2}$. For comparison, models of number counts in the *Herschel*/SPIRE bands suggest a space density of massive starburst galaxies at *z*>6 with $S_{500\mu m}$>30 mJy of 0.014 deg$^{-2}$ (ref. 7).

To understand the nature of galaxies selected by this technique, we have obtained full frequency scans of the 3-mm and 1-mm bands toward HFLS3 (also known as 1HERMES S350 J170647.8+584623; $S_{500\mu m}/S_{350\mu m}$ = 1.45), the brightest candidate discovered in our study. These observations, augmented by selected follow-up over a broader wavelength range, unambiguously determine the galaxy redshift to be *z*=6.3369+/-0.0009 based on a suite of 7 CO lines, 7 $H_2O$ lines, and OH, $OH^+$, $H_2O^+$, $NH_3$, [CI], and [CII] lines detected in emission and absorption (Figure 1). At this redshift, the Universe was just 880 million years old (or 1/16$^{th}$ of its present age), and 1" on the sky corresponds to a physical scale of 5.6 kpc. Further observations from optical to radio wavelengths reveal strong continuum emission over virtually the entire wavelength range between 2.2 μm and 20 cm, with no detected emission shortward of 1 μm (see Supplementary Information Section 2 and Figures S1-S11 for additional details).

HFLS3 hosts an intense starburst. The 870 μm-flux of HFLS3 is >3.5 times higher than those of the brightest high-redshift starbursts in a 0.25-deg$^2$ region containing the Hubble Ultra Deep Field (HUDF).[8] From the continuum spectral energy distribution (Fig. 2), we find that the far-infrared (FIR) luminosity $L_{FIR}$ and inferred star formation rate (SFR) of 2,900 $M_{sun}yr^{-1}$ of HFLS3 are 15-20 times those of the prototypical local ultra-luminous starburst Arp 220, and >2,000 times those of the Milky Way (Table 1 and Supplementary Information Section 3). The SFR of HFLS3 alone corresponds to ~4.5 times the ultraviolet-based SFR of all *z*=5.5-6.5 star-forming galaxies in the HUDF combined,[9] but the rarity and dust obscuration of ultra-red sources like HFLS3 implies that they do not dominate the UV photon density needed to reionize the universe.[10]

HFLS3 is a massive, gas-rich galaxy. From the spectral energy distribution and the intensity of the CO and [CII] emission, we find a dust mass of $M_d$=1.3 x 10$^9$ $M_{sun}$ and total molecular and atomic gas masses of $M_{gas}$=1.0 x 10$^{11}$ $M_{sun}$ and $M_{HI}$=2.0 x 10$^{10}$ $M_{sun}$. These masses are 15-20 times those of Arp 220, and correspond to a gas-to-dust ratio of ~80 and a gas depletion timescale of $M_{gas}$/SFR ~36 Myr. These values are comparable to lower-redshift submillimeter-selected starbursts.[11,12] From the [CI] luminosity, we find an atomic carbon mass of 4.5 x 10$^7$ $M_{sun}$. At the current star formation rate of HFLS3, this level of carbon enrichment could have been achieved through supernovae on a timescale of ~$10^7$ yr.[13] The profiles of the molecular and atomic emission lines typically show two velocity components (Figs. 1, S5, and S7). The gas is distributed over a 1.7 kpc radius region with a high velocity gradient and dispersion (Fig. 3). This suggests a dispersion-dominated galaxy with a dynamical mass of $M_{\rm dyn}$ = 2.7 x $10^{11}$ $M_{\rm sun}$. The gas mass fraction in galaxies is a measure of the relative depletion and replenishment of molecular gas, and is expected to be a function of halo mass and redshift from simulations.[14] In HFLS3, we find a high gas mass fraction of $f_{\rm gas} = M_{\rm gas}/M_{\rm dyn} \sim 40\%$, comparable to what is found in submillimeter-selected starbursts and massive star-forming galaxies at z~2,[15,16] but ~3 times higher than in nearby ultra-luminous infrared galaxies (ULIRGs) like Arp 220, and >30 times higher than in the Milky Way. From population synthesis modeling, we find a stellar mass of $M_*$ = 3.7 x $10^{10}$ $M_{\rm sun}$, comparable to that of Arp 220 and about half that of the Milky Way. This suggests that at most ~40% of $M_{\rm dyn}$ within the radius of the gas reservoir are due to dark matter. With up to ~$10^{11}$ $M_{\rm sun}$ of dark matter within 3.4 kpc, HFLS3 likely resides in a dark matter halo massive enough to grow a present-day galaxy cluster.[17] The efficiency for star formation is given by $\varepsilon=t_{\rm dyn}$ x SFR/$M_{\rm gas}$, where $t_{\rm dyn}=(r^3/(2GM))^{1/2}$ is the dynamical (or free-fall) time, $r$ is the source radius, $M$ is the mass within radius $r$ and $G$ is the gravitational constant. For $r$=1.7kpc and $M=M_{\rm gas}$, this suggests $\varepsilon$=0.06, which is a few times higher than found in nearby starbursts and in Giant Molecular Cloud cores in the Galaxy.[18]

The properties of the atomic and molecular gas in HFLS3 are fully consistent with a highly enriched, highly excited interstellar medium, as typically found in the nuclei of warm, intense starbursts, but distributed over a large, ~3.5-kpc-diameter region. The observed CO and [CII] luminosities suggest that dust is the primary coolant of the gas if both are thermally coupled. The $L_{\rm [CII]}/L_{\rm FIR}$ ratio of ~5 x $10^{-4}$ is typical for high radiation environments in extreme starbursts and active galactic nucleus (AGN) host galaxies.[19] The $L_{\rm [CII]}/L_{\rm CO(1\text{-}0)}$ ratio of ~3,000 suggests that the bulk of the line emission is associated with the photon dominated regions of a massive starburst. At the $L_{\rm FIR}$ of HFLS3, this suggests an infrared radiation field strength and gas density comparable to nearby ULIRGs without luminous AGN (Figs. 4 & 5 of ref. 19).

From the spectral energy distribution of HFLS3, we derive a dust temperature of $T_{\rm dust}=56^{+9}_{-12}$ K, ~10 K less than in Arp 220, but ~3 times that of the Milky Way. CO radiative transfer models assuming collisional excitation suggest a gas kinetic temperature of $T_{\rm kin}=144^{+59}_{-30}$ K and a gas density of $\log_{10}(n(H_2))=3.80^{+0.28}_{-0.17}$ cm$^{-3}$ (Supplementary Information Section 4 and Figures S13/S14). These models suggest similar gas densities as in nearby ULIRGs, and prefer $T_{\rm kin}>>T_{\rm dust}$, which may imply that the gas and dust are not in thermal equilibrium, and that the excitation of the molecular lines may be partially supported by the underlying infrared radiation field. This is consistent with the finding that we detect $H_2O$ and OH lines with upper level energies of $E/k_B$ > 300-450 K and critical densities of >$10^{8.5}$ cm$^{-3}$ at line intensities exceeding those of the CO lines. The intensities and ratios of the detected $H_2O$ lines cannot be reproduced by radiative transfer models assuming collisional excitation, but are consistent with being radiatively pumped by far-infrared photons, at levels comparable to those observed in Arp 220 (Figures S15/S16).[20,21] The CO and $H_2O$ excitation is inconsistent with what is observed in quasar host galaxies like Mrk 231 and APM08279+5255 at $z$=3.9, which lends support to the conclusion that the gas is excited by a mix of collisions and infrared photons associated with a massive, intense starburst, rather than hard radiation associated with a luminous AGN.[22] The physical conditions in the ISM of HFLS3 thus are comparable to those in the nuclei of the most extreme nearby starbursts, consistent with the finding that it follows the radio-FIR correlation for star-forming galaxies.

HFLS3 is rapidly assembling its stellar bulge through star formation at surface densities close to the theoretically predicted limit for "maximum starbursts".[23] At a rest-frame wavelength of 158 μm, the FIR emission is distributed over a relatively compact area of 2.6 kpc x 2.4 kpc physical diameter along its major and minor axes respectively (Fig. 3; as determined by elliptical Gaussian fitting). This suggests an extreme star formation rate surface density of $\Sigma_{\rm SFR}$~600 $M_{\rm sun}$yr$^{-1}$ kpc$^{-2}$ over a 1.3-kpc-radius region, and is consistent with near-Eddington-limited star formation if the starburst disk is supported by radiation pressure.[24] This suggests the presence of a kiloparsec-scale hyper-starburst similar to that found in the z=6.42 quasar

J1148+5251.[25] Such high $\Sigma_{SFR}$ are also observed in the nuclei of local ULIRGs such as Arp 220, albeit on two orders of magnitude smaller scales. A starburst at such high $\Sigma_{SFR}$ may produce strong winds. Indeed, the relative strength and broad, asymmetric profile of the OH $^2\Pi_{1/2}$(3/2-1/2) doublet detected in HFLS3 may indicate a molecular outflow, reminiscent of the OH outflow in Arp 220.[21]

The identification of HFLS3 alone is still consistent with the model-predicted space density of massive starburst galaxies at $z$>6 with $S_{500\mu m}$ > 30 mJy of 0.014 deg$^{-2}$ (ref. 7). This corresponds to only $10^{-3}$-$10^{-4}$ times the space density of Lyman-break galaxies at the same redshift, but is comparable to the space density of the most luminous quasars hosting supermassive black holes (SMBHs, i.e., a different population of massive galaxies) at such early cosmic times.[26] The host galaxies around these very distant SMBHs are commonly FIR-luminous, but less intensely star-forming, with typically a few times lower $L_{FIR}$ than ultra-red sources.[25] This highlights the difference between selecting massive z>6 galaxies at the peak of their star formation activity through $L_{FIR}$, and at the peak of their black hole activity through luminous AGN. The substantial population of ultra-red sources discovered with Herschel will be an ideal probe of early galaxy evolution and heavy element enrichment within the first billion years of cosmic time. These galaxies are unlikely to dominate the star formation history of the Universe at z>6,[5] but they trace the highest peaks in SFR at early epochs. A detailed study of this galaxy population will reveal the mass and redshift distribution, number density and likely environments of such objects, which if confirmed in larger numbers may present a stern challenge to current models of early cosmic structure formation.

## References:

1. Chapman, S.C. *et al.* A median redshift of 2.4 for galaxies bright at submillimetre wavelengths. *Nature* **422**, 695-698 (2003).
2. Capak, P. *et al.* A massive protocluster of galaxies at a redshift of z~5.3. *Nature* **470**, 233-235 (2011).
3. Walter, F. *et al.* The intense starburst HDF 850.1 in a galaxy overdensity at z ≈ 5.2 in the Hubble Deep Field. *Nature* **486**, 233-236 (2012).
4. Vieira, J.D. *et al.* Dusty starburst galaxies in the early Universe as revealed by gravitational lensing. *Nature*, in press.
5. Bouwens, R. *et al.* A candidate redshift z~10 galaxy and rapid changes in that population at an age of 500Myr. *Nature* **469**, 504-507 (2011).
6. Oliver, S. *et al.* The Herschel Multi-tiered Extragalactic Survey: HerMES. *Mon. Not. R. Astron. Soc.* **424**, 1614-1635 (2012).
7. Béthermin, M. *et al.* A Unified Empirical Model for Infrared Galaxy Counts Based on the Observed Physical Evolution of Distant Galaxies. *Astrophys. J. Lett.* **757**, L23 (2012).
8. Karim, A. *et al.* An ALMA survey of submillimetre galaxies in the Extended Chandra Deep Field South: High resolution 870μm source counts. Submitted to *Mon. Not. R. Astron. Soc.; arXiv* preprint at <http://arxiv.org/abs/1210.0249> (2012).
9. Bouwens, R.J. *et al.* Galaxies at z ~ 6: The UV Luminosity Function and Luminosity Density from 506 HUDF, HUDF Parallel ACS Field, and GOODS i-Dropouts. *Astrophys. J.* **653**, 53-85 (2006).
10. Robertson, B. *et al.* Early star-forming galaxies and the reionization of the Universe. *Nature* **468**, 49-55 (2010).
11. Michalowski, M.J. *et al.* Rapid Dust Production in Submillimeter Galaxies at z > 4? *Astrophys. J.* **712**, 942-950 (2010).
12. Riechers, D.A. *et al.* Extended Cold Molecular Gas Reservoirs in z ~= 3.4 Submillimeter Galaxies. *Astrophys. J. Lett.* **739**, L31 (2011).
13. Walter, F. *et al.* Molecular gas in the host galaxy of a quasar at redshift z = 6.42. *Nature* **424**, 406-408 (2003).
14. Lagos, C. Del P. *et al.* On the impact of empirical and theoretical star formation laws on galaxy formation. *Mon. Not. R. Astron. Soc.* **416**, 1566-1584 (2011).
15. Tacconi, L.J. *et al.* Submillimeter Galaxies at z ~ 2: Evidence for Major Mergers and Constraints on Lifetimes, IMF, and CO-$H_2$ Conversion Factor. *Astrophys. J.* **680**, 246-262 (2008).
16. Tacconi, L.J. *et al*. High molecular gas fractions in normal massive star-forming galaxies in the young Universe. *Nature* **463**, 781-784 (2010).
17. Overzier, R. *et al.* ΛCDM predictions for galaxy protoclusters - I. The relation between galaxies, protoclusters and quasars at z ~ 6. *Mon. Not. R. Astron. Soc.* **394**, 577-594 (2009).
18. Krumholz, M.R. *et al.* A Universal, Local Star Formation Law in Galactic Clouds, nearby Galaxies, High-redshift Disks, and Starbursts. *Astrophys. J.* **745**, 69 (2012).
19. Stacey, G.J. *et al.* A 158 μm [C II] Line Survey of Galaxies at z ~ 1-2: An Indicator of Star Formation in the Early Universe. *Astrophys. J.* **724**, 957-974 (2010).
20. Rangwala, N. *et al.* Observations of Arp 220 Using Herschel-SPIRE: An Unprecedented View of the Molecular Gas in an Extreme Star Formation Environment. *Astrophys. J.* **743**, 94 (2011).

21. Gonzalez-Alfonso, E. *et al.* Herschel/PACS spectroscopy of NGC 4418 and Arp 220: $H_2O$, $H_2{}^{18}O$, OH, $^{18}OH$, O I, HCN and $NH_3$. *Astron. Astrophys.* **541**, A4 (2012).
22. van der Werf, P. *et al.* Water Vapor Emission Reveals a Highly Obscured, Star-forming Nuclear Region in the QSO Host Galaxy APM 08279+5255 at z = 3.9. *Astrophys. J. Lett.* **741**, L38 (2011).
23. Elmegreen, B.G. Galactic Bulge Formation as a Maximum Intensity Starburst. *Astrophys. J.* **517**, 103-107 (1999).
24. Thompson, T. *et al.* Radiation Pressure-supported Starburst Disks and Active Galactic Nucleus Fueling. *Astrophys. J.* **630**, 167-185 (2005).
25. Walter, F. *et al.* A kiloparsec-scale hyper-starburst in a quasar host less than 1 gigayear after the Big Bang. *Nature* **457**, 699-701 (2009).
26. Jiang, L. *et al.* A Survey of z ~ 6 Quasars in the Sloan Digital Sky Survey Deep Stripe. II. Discovery of Six Quasars at $z_{AB}$>21. *Astronomical. J.* **138**, 305-311 (2009).
27. Downes, D. & Solomon, P.M. Rotating Nuclear Rings and Extreme Starbursts in Ultraluminous Galaxies. *Astrophys. J.* **507**, 615-654 (1998).
28. Sodroski, T.J. *et al.* Large-scale characteristics of interstellar dust from COBE DIRBE observations. *Astrophys. J.* **428**, 638-646 (1994).
29. Murray, N. & Rahman, M. Star Formation in Massive Clusters Via the Wilkinson Microwave Anisotropy Probe and the Spitzer Glimpse Survey. *Astrophys. J.* **709**, 424-435 (2010).
30. McMillan, P.J. Mass models of the Milky Way. *Mon. Not. R. Astron. Soc.* **414**, 2446-2457 (2011).

**Supplementary Information** is linked to the online version of the paper at www.nature.com/nature.

**Acknowledgments:** Herschel is an ESA space observatory with science instruments provided by European-led Principal Investigator consortia and with important participation from NASA. This research has made use of data from HerMES project (http://hermes.sussex.ac.uk/). HerMES is a Herschel Key Programme utilising Guaranteed Time from the SPIRE instrument team, ESAC scientists and a mission scientist. See Supplementary Information for further acknowledgments.

**Author Contributions:** D.R. had the overall lead of the project. C.M.B., D.C., I.P.-F., R.I., C.B., H.F., J.V., and R.N. have contributed significantly to the taking and analysis of the follow-up data with different instruments by leading several telescope proposals and analysis efforts. C.D.D. has led the selection of the parent sample. A.C., J.W., J.C., A.C., P.H., and J.K. have contributed significantly to the data analysis and to fitting and modeling the results. All other authors contributed to the proposals, source selection, data analysis and interpretation, in particular through work on the primary *Herschel* SPIRE data in which the source was discovered through the HerMES consortium (led by J.B. and S.O.). All authors have reviewed, discussed, and commented on the manuscript.

**Competing Interest Statement:** The authors declare that they have no competing financial interests.

**Corresponding authors:** Correspondence and requests for material should be addressed to Dominik Riechers (dr@astro.cornell.edu).

## Figures:

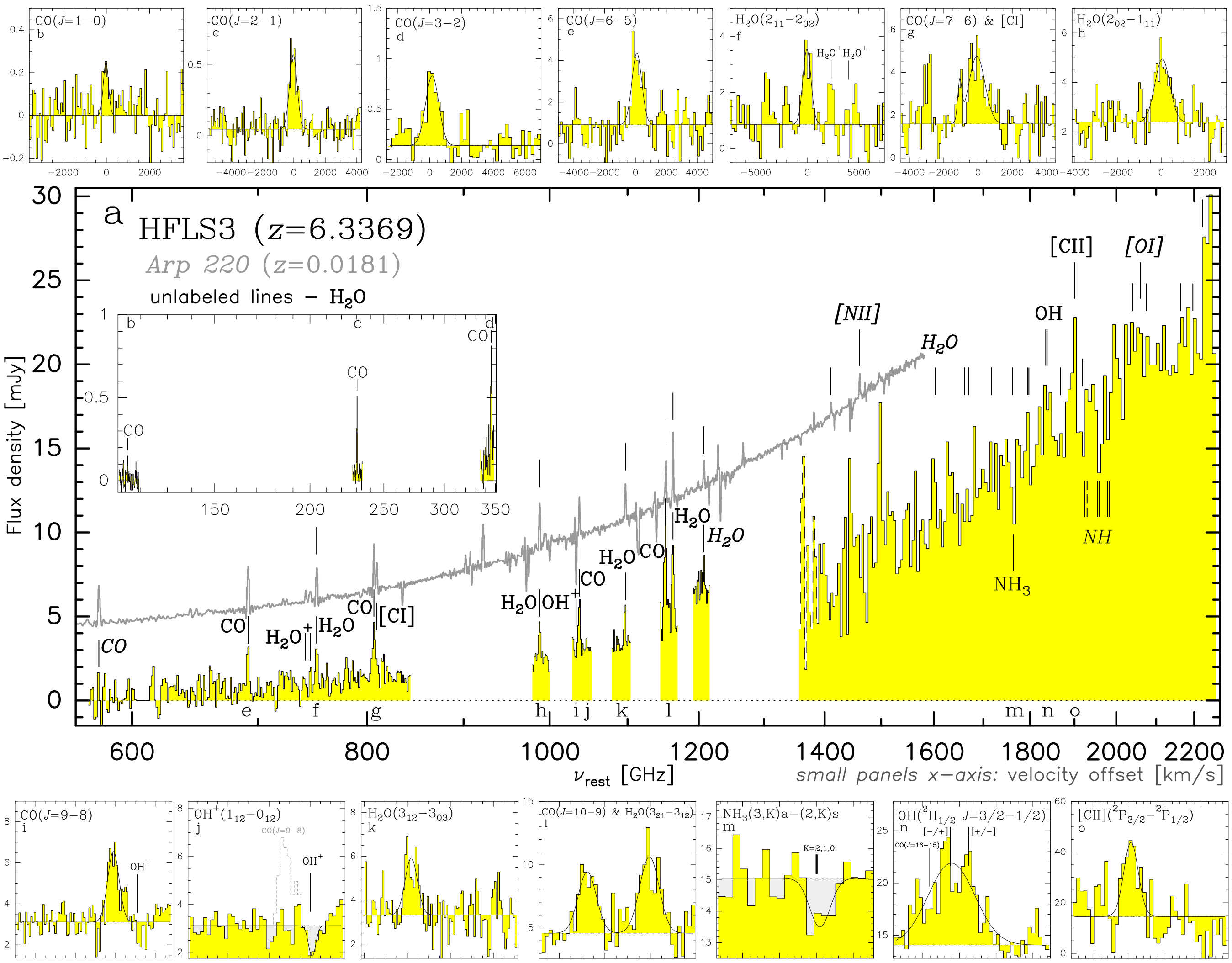

**Figure 1: Redshift identification through molecular and atomic spectroscopy of HFLS3. a**, Black trace, wide-band spectroscopy in the observed-frame 19 - 0.95-mm (histogram; rest-frame 2,600 - 130 μm) wavelength range with CARMA (3 mm; "blind" frequency scan of the full band), the PdBI (2 mm), the JVLA (19 - 6 mm), and CSO/Z-spec (1 mm; instantaneous coverage). (CARMA, Combined Array for Research in Millimeter-wave Astronomy; PdBI, Plateau de Bure Interferometer; JVLA, Jansky Very Large Array; and CSO, Caltech Submillimeter Observatory) This uniquely determines the redshift of HFLS3 to be *z*=6.3369 based on the detection of a series of $H_2O$, CO, OH, $OH^+$, $NH_3$, [CI] and [CII] emission and absorption lines. **b** to **o**, Detailed profiles of detected lines (histograms; rest frequencies are indicated by corresponding letters in **a**). 1-mm lines (**m**-**o**) are deeper, interferometric confirmation observations for $NH_3$, OH (both PdBI), and [CII] (CARMA) not shown in **a**. The line profiles are typically asymmetric relative to single Gaussian fits, indicating the presence of two principal velocity components at redshifts of 6.3335 and 6.3427. The implied CO, [CI], and [CII] line luminosities are 5.08+/-0.45 x $10^6$, 3.0+/-1.9 x $10^8$, and 1.55+/-0.32 x $10^{10}$ $L_{sun}$. Strong rest-frame submillimeter to far-infrared continuum emission is detected over virtually the entire wavelength range. For comparison, the *Herschel*/SPIRE spectrum of the nearby ultra-luminous infrared galaxy Arp 220[20] is overplotted in grey (**a**). Lines labeled in italic are tentative detections or upper limits (see Table S2). Most of the bright spectral features detected in Arp 220[20,21] are also detected in HFLS3 (in spectral regions not blocked by the terrestrial atmosphere). See Supplementary Information Sections 2-4 for more details.

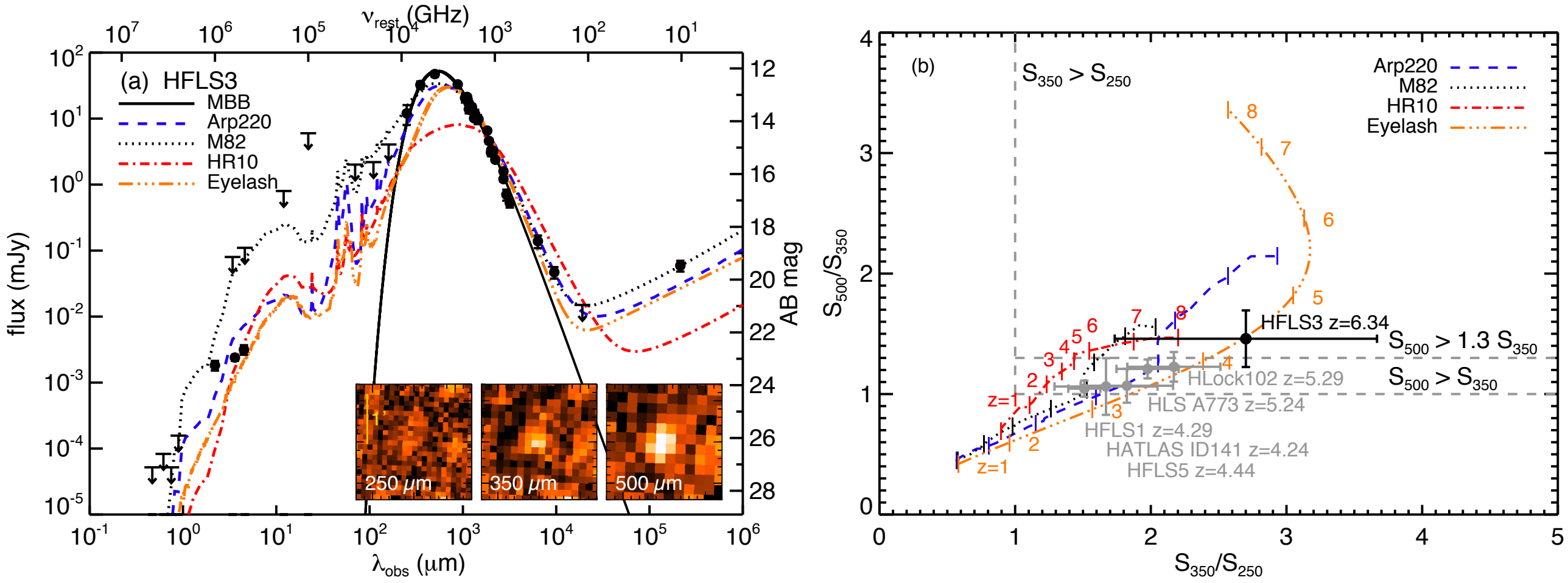

**Figure 2: Spectral energy distribution (SED) and *Herschel*/SPIRE colors of HFLS3. a**, HFLS3 was identified as a very high redshift candidate, as it appears red between the *Herschel*/SPIRE 250-, 350-, and 500-μm bands (inset). The SED of the source (data points; $\lambda_{obs}$, observed-frame wavelength; $\nu_{rest}$, rest-frame frequency; AB mag, magnitudes in the AB system; error bars are 1$\sigma$ r.m.s. uncertainties in both panels) is fitted with a modified black body (MBB; solid line) and spectral templates for the starburst galaxies Arp 220, M82, HR10, and the Eyelash (broken lines, see key). The implied FIR luminosity is $2.86^{+0.32}_{-0.31}$ x $10^{13}$ $L_{sun}$. The dust in HFLS3 is not optically thick at wavelengths longward of rest-frame 162.7 μm (95.4% confidence; Figure S12). This is in contrast to Arp 220, in which the dust becomes optically thick (i.e., $\tau_d$=1) shortward of 234+/-3 μm.[20] Other high-redshift massive starburst galaxies (including the Eyelash) typically become optically thick around ~200 μm. This suggests that none of the detected molecular/fine structure emission lines in HFLS3 require correction for extinction. The radio continuum luminosity of HFLS3 is consistent with the radio-FIR correlation for nearby star-forming galaxies. **b**, 350 μm/250 μm and 500 μm/350 μm flux density ratios of HFLS3. The colored lines are the same templates as in **a**, but redshifted between 1<$z$<8 (number labels indicate redshifts). Dashed grey lines indicate the dividing lines for red ($S_{250\mu m}<S_{350\mu m}<S_{500\mu m}$) and ultra-red sources ($S_{250\mu m}<S_{350\mu m}$ and 1.3 x $S_{350\mu m}$ < $S_{500\mu m}$). Gray symbols show the positions of five spectroscopically confirmed red sources at 4<z<5.5 (including three new sources from our study), which all fall outside the ultra-red cutoff. This shows that ultra-red sources will lie at $z$>6 for typical SED shapes (except those with low dust temperatures), while red sources typically are at z<5.5. See Supplementary Information Sections 1 and 3 for more details.

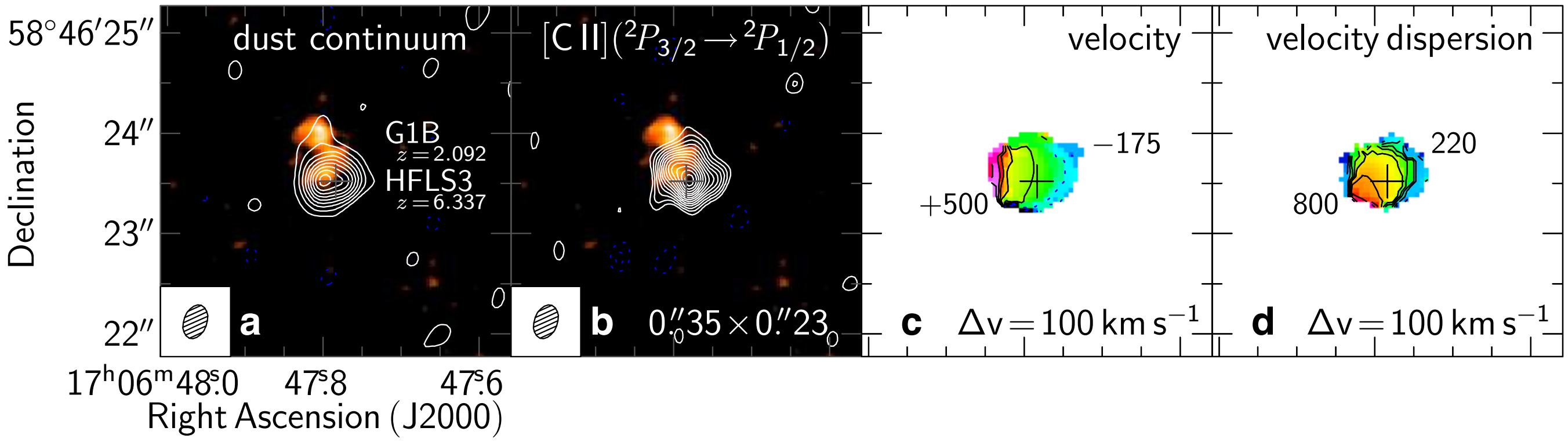

**Figure 3: Gas dynamics, dust obscuration, and distribution of gas and star formation in HFLS3.** **a**, **b**, High-resolution (FWHM 0.35"x0.23") maps of the 158-μm continuum (**a**) and [CII] line emission (**b**) obtained at 1.16 mm with the PdBI in A-configuration, overlaid on a Keck/NIRC2 2.2-μm adaptive optics image (rest-frame UV/optical light). The r.m.s. uncertainty in the continuum (**a**) and line (**b**) maps is 180 and 400 μJy beam$^{-1}$, and contours are shown in steps of 3 and 1$\sigma$, starting at 5 and 3$\sigma$, respectively. A $z$=2.092 galaxy (labeled G1B) identified through Keck/LRIS spectroscopy is detected ~0.65" north of HFLS3, but is not massive enough to cause significant gravitational lensing at the position of HFLS3. Faint infrared emission is detected toward a region with lower dust obscuration in the north-eastern part of HFLS3 (not detected at <1 μm). The Gaussian diameters of the resolved [CII] and continuum emission are 3.4 kpc x 2.9 kpc and 2.6 kpc x 2.4 kpc, suggesting gas and SFR surface densities of $\Sigma_{gas}$ = 1.4 x 10$^4$ M$_{sun}$ pc$^{-2}$ and $\Sigma_{SFR}$ = 600 M$_{sun}$ yr$^{-1}$ kpc$^{-2}$ (~0.6 x 10$^{13}$ L$_{sun}$ kpc$^{-2}$). The high $\Sigma_{SFR}$ is consistent with a maximum starburst at near-Eddington-limited intensity. Given the moderate optical depth of $\tau_d$<~1 at 158 μm, this estimate is somewhat conservative. Peak velocity (**c**) and F.W.H.M. velocity dispersion (**d**) maps of the [CII] emission are obtained by Gaussian fitting to the line emission in each spatial point of the map. Velocity contours are shown in steps of 100 kms$^{-1}$. High-resolution CO $J$=7-6 and 10-9 and $H_2O$ $3_{21}$-$3_{12}$ observations show consistent velocity profiles and velocity structure (Figures S5-S7). The large velocity dispersion suggests that the gas dynamics in this system are dispersion-dominated. See Supplementary Information Sections 3 and 5 for more details.

**Table 1: Observed and derived quantities for HFLS3, Arp 220 and the Galaxy**

| | HFLS3 | Arp 220* | Milky Way* |
|---|---|---|---|
| **redshift** | 6.3369 | 0.0181 | - |
| $M_{gas}$ (M$_{sun}$)[a] | (1.04+/-0.09) x 10$^{11}$ | 5.2 x 10$^9$ | 2.5 x 10$^9$ |
| $M_{dust}$ (M$_{sun}$)[b] | $1.31^{+0.32}_{-0.30}$ x 10$^9$ | ~1 x 10$^8$ | ~6 x 10$^7$ |
| $M_*$ (M$_{sun}$)[c] | ~3.7 x 10$^{10}$ | ~3-5 x 10$^{10}$ | ~6.4 x 10$^{10}$ |
| $M_{dyn}$ (M$_{sun}$)[d] | 2.7 x 10$^{11}$ | 3.45 x 10$^{10}$ | 2 x 10$^{11}$ (<20 kpc) |
| $f_{gas}$[e] | 40% | 15% | 1.2% |
| $L_{FIR}$ (L$_{sun}$)[f] | $2.86^{+0.32}_{-0.31}$ x 10$^{13}$ | 1.8 x 10$^{12}$ | 1.1 x 10$^{10}$ |
| **SFR** (M$_{sun}$yr$^{-1}$)[g] | 2,900 | ~180 | 1.3 |
| $T_{dust}$ (K)[h] | $55.9^{+9.3}_{-12.0}$ | 66 | ~19 |

For details see Supplementary Information, Section 3.

[*]Literature values for Arp 220 and the Milky Way are adopted from refs. 27, 20, 28, 29, and 30. The total molecular gas mass of the Milky Way is uncertain by at least a factor of 2. Quoted dust masses and stellar masses are typically uncertain by factors of 2-3 due to systematics. The dynamical mass for the Milky Way is quoted within the inner 20 kpc to be comparable to the other systems, not probing the outer regions dominated by dark matter. The dust temperature in the Milky Way varies by at least +/-5 K around the quoted value, which is used as a representative value. Both Arp 220 and the Milky Way are known to contain small fractions of significantly warmer dust. All error bars are 1$\sigma$ r.m.s. uncertainties.

[a]Molecular gas mass, derived assuming $\alpha_{CO} = M_{gas}/L'_{CO}$ = 1 M$_{sun}$ (K kms$^{-1}$pc$^2$)$^{-1}$, see Supplementary Information, Section 3.3.

[b]Dust mass, derived from spectral energy distribution fitting, see Supplementary Information, Section 3.1.

[c]Stellar mass, derived from population synthesis fitting, see Supplementary Information, Section 3.4.

[d]Dynamical mass, see Supplementary Information, Section 3.5.

[e]Gas mass fraction, derived assuming $f_{gas}=M_{gas}/M_{dyn}$, see Supplementary Information, Section 3.6.

[f]Far-infrared luminosity as determined over the range of 42.5-122.5 μm from spectral energy distribution fitting, see Supplementary Information, Section 3.1.

[g]Star formation rate, derived assuming SFR[M$_{sun}$yr$^{-1}$] = 1.0 x 10$^{-10}$ L$_{FIR}$ [L$_{sun}$], see Supplementary Information, Section 3.2.

[h]Dust temperature, derived from spectral energy distribution fitting, see Supplementary Information, Section 3.1.

# 1. Target Selection and Source Densities

## 1.1 Selection of "Ultra-Red" Sources

Massive starburst galaxies at high redshift (z>1) form stars at very high rates, exceeding 1,000 $M_{sun}yr^{-1}$. The high star formation rates are usually supported by major mergers of gas-rich galaxies.[31,32] The starbursts are heavily enshrouded in the dust and gas that constitutes their mass reservoir for the formation of new stars. This material commonly absorbs the bulk of the rest-frame ultraviolet/optical light emitted by newly formed stars, making them faint at optical wavelengths. The absorbed light is re-emitted at rest-frame far-infrared wavelengths, creating a strong far-infrared bump in the spectral energy distribution.[31] The most distant of these galaxies thus are very luminous in the observed-frame (sub)millimeter continuum and molecular/atomic fine structure lines, which represent the main pathways of cooling in warm, dense clouds in their metal-enriched ISM.[33]

Optical/near-infrared color selection techniques, commonly employed to isolate very distant galaxies in large quantities, have revealed individual examples of massive starbursts out to $z$=5.3.[2,34] However, dust extinction leads to an effective bias against the systematic selection of such galaxies at these wavelengths.[3] Surveys of dusty galaxies at long submillimeter (>850 μm) to millimeter wavelengths probe the Rayleigh-Jeans part of the greybody dust emission spectrum over a large range in redshift, making them broadly insensitive to redshift.[31] Thus, neither of these techniques is suitable for a clean, systematic redshift selection of the most distant massive starbursts. However, sensitive, large area surveys in the short submillimeter at 250, 350, and 500 μm, as presently carried out with the *Herschel*/SPIRE instrument, typically probe the peak of the dust emission spectrum in high-redshift galaxies, making it possible to construct a submillimeter color-based redshift selection technique for distant starbursts based on the far-infrared dust bump. Dusty galaxies at $z$>3.5 will appear increasingly red in 250−500 μm colors, as these bands will increasingly probe the Wien part of the dust emission spectrum.

We have used this fact to search ~21 deg² of the *Herschel*/SPIRE data of the HerMES blank field survey[6] at 250, 350, and 500 μm wavelengths in the First Look Survey (FLS), Lockman Hole, and GOODS-North regions for "ultra-red" sources, i.e., galaxies that are significantly redder than the massive starbursts discovered thus far.[35,36] A systematic analysis of the SPIRE maps yields 44 solid candidate red sources with $S_{250\mu m}<S_{350\mu m}<S_{500\mu m}$, suggesting a source density of ≤2 $deg^{-2}$ down to a flux limit of 30 mJy at 500 μm (>5σ and above the confusion noise). Based on their colors, most of these sources are expected to be at high redshift, but only a few are expected to be associated with the most distant massive starbursts at $z$>6 (assuming similar SED shapes as starburst galaxies at low and high redshifts; see Figure 2). To isolate sources at the highest redshifts, we thus further refined our selection criteria within the candidate red source sample with $S_{500\mu m}/S_{350\mu m}$>1.3 (HFLS3: $S_{500\mu m}/S_{350\mu m}$ = 1.45). After the rejection of galaxies with non-thermal spectra and blended systems based on longer wavelength (sub)millimeter and radio continuum interferometry, this selection is expected to yield a fairly clean sample of rare galaxies with either very low dust temperatures or dusty starburst galaxies at very high redshift. Over the entire ~21 deg², this selection yields 5 candidate ultra-red sources (there are two more tentative identifications which however do not match the selection criteria for some flux extraction methods), corresponding to a source density of ≤0.24 $deg^{-2}$. From this sample, HFLS3 was isolated as the most promising z>6 galaxy candidate based on follow-up photometry at longer wavelengths.

## 1.2 Information on "Red" Sources

Two sources in the literature would be classified as red sources by our selections (see Fig. 2). The first source, H-ATLAS ID141, has 250, 350, and 500 μm fluxes of 115±19, 192±30, and 204±32 mJy.[35] Based on the detection of CO $J$=4−3 and 5−4 emission and the lower [CI] fine structure line redshifted to the 3 mm band with the IRAM PdBI, its redshift was determined to be z=4.243 (subsequently confirmed through the detection of additional lines).[35] The second source, HLS A773, has 250, 350, and 500 μm fluxes of 85±8, 168±8, and 203±9 mJy.[36] Based on the detection of CO $J$=5–4 and 6–5 emission redshifted to the 3 mm band with the IRAM 30m telescope, its redshift was determined to be 5.243 (subsequently confirmed through the detection of additional lines).[36] Both these sources are strongly gravitationally lensed, and neither of them fulfills the "ultra-red" selection criterion. As part of our survey for high redshift massive starburst galaxies, we have discovered three more red sources that do not fulfill the "ultra-red" criterion. Based on the detection of CO $J$=4–3 and 5–4 emission redshifted to the 3 mm band with CARMA, the redshift of HFLS1 was determined to be z=4.29. Based on the detection of CO $J$=4–3 and 5–4 emission and the lower [CI] fine structure line redshifted to the 3 mm band with CARMA, the redshift of HFLS5 was determined to be z=4.44. Based on the detection of CO $J$=5–4 and 6–5 emission redshifted to 3 mm with CARMA and the 205 μm [NII] line redshifted to 1 mm with the IRAM PdBI, the redshift of Lock-102 was determined to be z=5.29. Details on the properties of these sources will be reported in a subsequent publication. These sources are included in Fig. 2 for comparison to HFLS3.

## 2. Observations

### 2.1 *Herschel Space Observatory*

#### 2.1.1 Spectral and Photometric Imaging Receiver (SPIRE)

HFLS3 was observed with the SPIRE instrument[37] on board the *Herschel Space Observatory*[38] as part of the HerMES survey of the Extragalactic First Look Survey (XFLS) field (PIs: Bock, Oliver).[6] In the 250, 350 and 500 μm bands, the r.m.s. sensitivity at the position of HFLS3 is 2.3, 2.3, and 2.8 mJy, respectively (source extracted by fitting over multiple pixels, using an error map based on the rms of the bolometer readouts; the raw instrumental noise corresponds to 7.2, 7.3, and 8.3 mJy, respectively). These fitting errors do not account for confusion noise, which is at least ~6 mJy (1$\sigma$) in all SPIRE bands (accounted for in all subsequent analysis). The absolute flux scale is accurate to 5% (see Table S1 for continuum fluxes).

#### 2.1.2 Photodetector Array Camera and Spectrometer (PACS)

HFLS3 was observed with the PACS instrument[39] on board the *Herschel Space Observatory* on 2013 January 1 (PI: Riechers). Observations were carried out for a total of 3.9 hr in mini-scan mapping mode (4×15 repeats), using the 70+160 μm parallel mode and the 110+160 μm parallel mode for one orthogonal cross scan pair each. In the 70, 110 and 160 μm bands, the r.m.s. sensitivity at the position of HFLS3 is 0.67, 0.73, and 1.35 mJy, respectively. The absolute flux scale is accurate to 5% (see Table S1 for continuum fluxes).

### 2.2 Combined Array for Research in Millimeter-wave Astronomy (CARMA)

We used CARMA to scan the 77–112 GHz frequency range towards HFLS3 in an attempt to detect molecular emission lines to determine its redshift (PI: Riechers). For this "blind" CO scan at 3 mm, we used the wide-band correlator with 3.708 GHz of bandwidth per sideband at 5.208 MHz (15.6 kms$^{-1}$ at 100 GHz) resolution. We initially used two pairs of frequency setups, where the lower sideband of the second setup in each pair is used to fill in the gap in frequency coverage between the sidebands of the first setup. This allows us to cover almost the full frequency range with four tunings. Three additional setups were used to specifically target the regions covering tentatively identified lines in the initial scan, leading to higher sensitivity in these regions. A total of 20 observing runs were carried out between 2010 September 08 and 2011 July 14 under good 3 mm weather conditions in the compact D and E array configurations, yielding 46.0 hr (77.0 hr) time on source (total) after rejection of bad data, with a typical synthesized beam size of 4.5"×4.0" (natural baseline weighting).

Upon successful detection of redshifted CO $J$ = 6–5 and 7–6, [CI] $^3P_2$–$^3P_1$, and $H_2O$ $J_{\rm KaKc}$ = $2_{11}$–$2_{02}$ emission at $z$=6.3369 towards HFLS3 (rest-frame 691.4730763, 806.651806, 809.3435, and 752.0332 GHz), we targeted the [CII] $^2P_{3/2}$–$^2P_{1/2}$ fine structure line at rest-frame 1,900.543 GHz. We tuned the 1 mm receivers to 259.106 GHz in the upper sideband (USB), using the same correlator setup as above. A total of three observing runs were carried out between 2011 May 26 and June 04 under good 1 mm weather conditions in the D array configuration, yielding 4.3 hr (8.2 hr) time on source (total) after rejection of bad data, with a synthesized beam size of 2.0"×1.6" (natural baseline weighting).

For primary flux calibration, we observed Mars or Neptune. Passband calibration was obtained on the bright radio quasars 3C273, 3C279, 3C345, 3C454.3, J1635+381, J1733−130, and J1751+096. The nearby quasars 1824+568 or 3C345 were observed every 12−20 minutes for complex gain calibration.

The MIRIAD package was used for data editing, calibration, and imaging. In addition to detecting CO $J$ = 6−5 and 7−6, [CI] $^3P_2$−$^3P_1$, $H_2O$ $J_{KaKc}$ = $2_{11}$−$2_{02}$, and [CII] $^2P_{3/2}$−$^2P_{1/2}$ emission, we also tentatively detected $H_2O^+$ $J_{KaKc}$ = $2_{02}$−$2_{11}$ $J$ = 3/2−3/2 emission (rest-frame 746.1938 GHz), and obtained upper limits on the CO $J$ = 5−4 and 16−15 and $H_2O^+$ $J_{KaKc}$ = $2_{02}$-$2_{11}$ $J$ = 5/2−3/2 emission lines (rest-frame 576.2679305, 1,841.3455060, and 742.0332 GHz; see Table S2 for line fluxes and limits, and Figures 1 and S1 for spectra). Continuum emission is detected over the entire wavelength range (see Table S1 for continuum fluxes, and Figure S2 for continuum maps). Line-averaged maps for the CO $J$ = 6−5 and 7−6, [CI] $^3P_2$−$^3P_1$, $H_2O$ $J_{KaKc}$ = $2_{11}$−$2_{02}$, and [CII] $^2P_{3/2}$−$^2P_{1/2}$ lines were created using natural baseline weighting. The r.m.s. noise levels at the line frequencies are 0.25, 0.34, 0.40, 0.29, and 2.9 mJy beam$^{-1}$ per 994, 795, 566, 853, and 844 kms$^{-1}$ (312.5, 291.7, 208.3, 291.7, and 729.2 MHz) bin (see Figure S3 for line maps and synthesized beam sizes).

## 2.3 Caltech Submillimeter Observatory (CSO) Z-spec

We used the Z-spec single-beam grating spectrometer[40,41] on the CSO to observe the 190−308 GHz frequency range towards HFLS3 (PI: Bradford). Z-spec has a resolving power of ~1 part in 250−300 over the covered frequency range. We used the chop-and-nod mode with a 20s nod period and a 1.6 Hz, 90" chop. Observations were carried out between 2010 March 18 and May 01 under good 1 mm observing conditions ($\tau_{225GHz}$=0.039−0.197, with a mean of 0.090 and a median of 0.078), yielding 22.7 hr (39.5 hr) time on source (total) after rejection of bad data. This resulted in a typical sensitivity of 700 mJy s$^{1/2}$, corresponding to a per-channel r.m.s. of 2.6 mJy (40%−50% higher at the band edges, and with greater degradation below 200 GHz). The absolute calibration was carried out relative to planets, with quasars as secondary calibrators.

We detected the [CII] $^2P_{3/2}$−$^2P_{1/2}$ emission line at a signal-to-noise ratio of 3.4 (see Figures 1 and S4). We also tentatively detected the OH $^2\Pi_{1/2}$ 3/2−1/2 and $H_2O$ $J_{KaKc}$ = $4_{13}$−$4_{04}$ emission lines at signal-to-noise ratios of 1.8 and 1.5, and the $NH_3$ 3s−2a and NH $2_2$−$1_1$ absorption features at signal-to-noise ratios of 2.0 and 2.3. In addition, these data provide limits on the CO $J$=12−11 to 19−18 lines, as well as on [NII] $^3P_1$−$^3P_0$, [OI] $^3P_1$−$^3P_0$, CH 2,3/2−1,1/2, and $H_2O$ $J_{KaKc}$ = $5_{23}$−$5_{14}$ to $5_{14}$−$5_{05}$ (see Table S2). Continuum emission with a steeply rising slope is detected over the entire wavelength range (see Table S1 for continuum fluxes obtained after spectral averaging into 5 bins).

## 2.4 Plateau de Bure Interferometer (PdBI)

We used the IRAM PdBI with 4−6 antennas in the compact C and D configurations to target HFLS3 in 10 frequency settings at 3 mm, 2 mm, and 1 mm to observe 15 redshifted emission/absorption lines from the CO, $H_2O$, OH, $OH^+$, $CH^+$, and $NH_3$ molecules (PIs: Riechers, Perez-Fournon). Observations were carried out between 2011 August 12 and December 23 and 2012 May 24−31. We also used the PdBI with 5−6 antennas in the most extended A configuration to target HFLS3 in 3 frequency settings at 3 mm, 2 mm, and 1 mm, to image the CO $J$ = 7−6 and 10−9, [CI] $^3P_2$−$^3P_1$, $H_2O$ $J_{KaKc}$ = $3_{21}$−$3_{12}$, and [CII] $^2P_{3/2}$−$^2P_{1/2}$ emission lines (2 mm setting also observed in D configuration). These observations were carried out between 2012 February 3 and 24. All observations used the WideX correlator with a total bandwidth of 3.6 GHz at a spectral resolution of 2 MHz (dual polarization). The 6 antenna-equivalent on source times for our twelve frequency setups tuned to 110.128 (CO $J$ = 7−6 and [CI] $^3P_2$−$^3P_1$), 113.819 ($CH^+$ $J$ = 1−0), 134.650 ($H_2O$ $J_{KaKc}$ = $2_{02}$−$1_{11}$), 141.328 (CO $J$=9−8 and $OH^+$ $1_{1F}$−$0_{1F}$), 148.800 ($H_2O$ $J_{KaKc}$ = $3_{12}$−$3_{03}$), 157.757 (CO $J$=10−9 and $H_2O$ $J_{KaKc}$ = $3_{21}$−$3_{12}$ and $3_{12}$−$2_{21}$), 164.598 ($H_2O$ $J_{KaKc}$ = $4_{22}$−$4_{13}$), 204.027 (CO $J$=13−12), 227.604 ($H_2O$ $J_{KaKc}$ = $2_{12}$−$1_{01}$ and $2_{21}$−$2_{12}$), 240.364 ($NH_3$ 3Ka−2Ks), 250.490 (OH $^2\Pi_{1/2}$ 3/2−1/2 and CO $J$=16−15), and 259.039 GHz ([CII] $^2P_{3/2}$−$^2P_{1/2}$) are 5.0, 10.2, 4.3, 6.7, 2.4, 7.8, 1.8, 1.1, 1.9, 3.0, 2.3, and 4.5 hr, respectively. The nearby radio quasars 1637+574, 1849+670, or 1642+690 were observed every 22.5 minutes for complex gain and bandpass calibration. Several regularly monitored millimeter sources were observed during each track for absolute flux calibration, yielding a flux scale that is accurate to 5%–10%.

For data reduction, calibration, and imaging, the GILDAS package was used. We solidly detect the CO $J$ = 9–8 and 10–9, OH $^2\Pi_{1/2}$ 3/2–1/2, and $H_2O$ $J_{KaKc}$ = $2_{02}$–$1_{11}$, $3_{12}$–$3_{03}$, $3_{12}$–$2_{21}$, and $3_{21}$–$3_{12}$ emission lines, and tentatively detect the CO $J$=13–12 and $H_2O$ $J_{KaKc}$ = $4_{22}$–$4_{13}$ emission and $OH^+$ $1_{1F}$–$0_{1F}$ and $NH_3$ 3Ka–2Ks absorption lines. We additionally obtain upper limits on the $CH^+$ $J$ = 1–0, CO $J$ = 16–15, and $H_2O$ $J_{KaKc}$ = $2_{12}$–$1_{01}$ and $2_{21}$–$2_{12}$ lines (see Figure 1 for line detections, Figure S1 for spectra of tentatively detected lines and upper limits, and Table S2 for line fluxes). Continuum emission is detected over the entire wavelength range (see Table S1 for continuum fluxes, and Figure S2 for continuum maps). We spatially resolve the CO $J$ = 7–6 and 10–9, [CI] $^3P_2$–$^3P_1$, $H_2O$ $J_{KaKc}$ = $3_{21}$–$3_{12}$, and [CII] $^2P_{3/2}$–$^2P_{1/2}$ emission lines and underlying continuum emission in the most extended array configuration. Line-averaged maps of the compact configuration data for the CO $J$ = 9–8 and 10–9, OH $^2\Pi_{1/2}$ 3/2–1/2, and $H_2O$ $J_{KaKc}$ = $2_{02}$–$1_{11}$, $3_{12}$–$3_{03}$, and $3_{21}$–$3_{12}$ lines were created using natural baseline weighting. The r.m.s. noise levels at the line frequencies are 0.22, 0.46, 0.32, 0.25, 0.30, and 0.48 mJy beam$^{-1}$ per 1,527, 993, 2,300, 802, 802, and 1,059 kms$^{-1}$ (720, 520, 1,920, 360, 400, and 560 MHz) bin (see Figure S3 for line maps and synthesized beam sizes). High-resolution line-averaged maps of the CO $J$ = 7–6 and 10–9, $H_2O$ $J_{KaKc}$ = $3_{21}$–$3_{12}$, and [CII] $^2P_{3/2}$–$^2P_{1/2}$ emission lines were created using uniform baseline weighting, and a high-resolution [CI] $^3P_2$–$^3P_1$ map was created using natural weighting. The r.m.s. noise levels at the line frequencies are 0.16, 0.11, 0.12, 0.40, and 0.19 mJy beam$^{-1}$ per 982, 1,184, 984, 625, and 652 kms$^{-1}$ (360, 620, 520, 540, and 240 MHz) bin (see Figures 3 and S5 for line maps and synthesized beam sizes). High-resolution velocity and dispersion maps were created for the CO and [CII] lines by blanking the data below 1σ level (Figures 3 and S6). Separate [CII] spectra were extracted for the two main subcomponents of the galaxy (Figure S7). High-resolution continuum maps at 2.7, 1.9, and 1.2 mm were created using uniform baseline weighting. The r.m.s. noise levels over the line-free regions (2,740, 1,960, and 2,800 MHz) are 64, 63, and 150 μJy beam$^{-1}$ (see Figure S2 for continuum maps and synthesized beam sizes).

## 2.5 Jansky VLA (JVLA)

We used the JVLA in the A array configuration to observe HFLS3 at 1.4 GHz (PI: Ivison). Observations were taken on 2011 June 20 and 24 for a total of 4.0 hr, using the WIDAR correlator with a total bandwidth of 256 MHz (full polarization). One of the baseband pairs was lost during the first run, resulting in a 50% loss in effective bandwidth. Accounting for this loss, the total effective on-source integration time amounts to ~2.5 hr. The quasar 3C286 was observed as primary flux calibrator. The quasar J1643+6245 was observed every 30–40 minutes for secondary gain calibration. Data reduction, calibration and wide-field imaging were performed using the AIPS package, and following procedures outlined in the literature.[42,43] The synthesized beam size for these observations is 1.4”×1.2” (natural baseline weighting) at an r.m.s. noise level of 11 μJy beam$^{-1}$ over the full bandpass (see Table S1 for continuum flux and Figure S8 for continuum map).

We also used the JVLA in the DnC, D, and C array configurations to observe HFLS3 in the redshifted CO $J$=1–0, 2–1, and 3–2 emission lines (rest frequencies: 115.2712, 230.5380 and 345.7960 GHz, redshifted to 15.7112, 31.4217 and 47.1311 GHz; PIs: Riechers, Ivison). CO $J$=2–1 observations were taken under good weather conditions for 1.6 hr (2.5 hr) on source (total) on 2012 January 1, using the WIDAR correlator in *Ka* band with a total bandwidth of 2 GHz (full polarization) at a spectral resolution of 2 MHz (19 kms$^{-1}$). CO $J$=1–0 observations were taken under good weather conditions for 0.9 hr (1.5 hr) on source (total) on 2012 January 12, using the same correlator setup in *U* band, yielding a spectral resolution of 38 kms$^{-1}$. CO $J$=3–2 observations were taken under good weather conditions for 2.8 hr (4.5 hr) on source (total) on 2012 April 15 and 16, using the same correlator setup in *Q* band, yielding a spectral resolution of 12.7 kms$^{-1}$. The quasar 3C286 was observed as primary flux calibrator. The quasar J1638+5720 was observed every 4.5 minutes for secondary gain calibration. Data reduction and calibration were performed using the AIPS package. The synthesized beam sizes for the CO $J$=3–2, 2–1 and 1–0 observations are 0.65”×0.51”, 2.9”×2.3”, and 5.3”×3.4” (natural baseline weighting) at r.m.s. noise levels of 93, 31 and 42 μJy beam$^{-1}$ per 1,183, 763, and 611 kms$^{-1}$ bin (corresponding to 186, 80, and 32 MHz; see Table S2 for line parameters and Figures 1 and S3 for CO line spectra and CO $J$=2−1 map).

## 2.6 Submillimeter Array (SMA)

We used the SMA to observe HFLS3 at 890 μm and 1.1 mm. Observations in the compact (COM) and subcompact (SUB) configurations at 270/275 GHz (local oscillator frequency) were carried out for 7.1 hr on source on 2010 May 12 and 2011 Mar 14/23 with six or seven antennas (PI: Clements). The photometry and astrometry from the initial observations in 2010 were used in combination with the SPIRE photometry to refine the photometric redshift and position for all subsequent follow-up. Further observations were carried out in the extended (EXT) configuration for 1.3 hr on source at 270 GHz on 2011 August 02 with eight antennas. We also observed the source in the SUB configuration for 4.7 hr on source at 341 GHz on 2011 July 06 with seven antennas, in the EXT configuration for 5.2 hr on source on 2011 August 03 with eight antennas, and in the very extended (VEX) configuration at 341 GHz for 3.8 hr on source on 2011 September 06 with eight antennas. In all tracks, the signal was averaged over both sidebands (4 GHz of bandwidth each in the 4−8 GHz intermediate frequency range) to maximize the effective continuum sensitivity. Flux, bandpass, and gain calibration were performed on nearby quasars. In compact configurations, Callisto was used for absolute flux calibration. Observations of a test quasar in the VEX configuration suggests that the absolute astrometry after accounting for phase noise is accurate to <0.1".

Calibration was performed using the MIR package, and the visibility data were imaged using the AIPS package (see Table S1 for continuum fluxes). Maps created at 1.1 mm yield resolutions of 3.4"×2.4" and 1.06"×0.97" in the COM and EXT configurations. Imaging of the VEX data at 890 μm and the EXT data at 1.1 mm yields resolutions of 0.36"×0.29" and 1.06"×0.97", respectively (maps created with natural baseline weighting). To maximize sensitivity and dynamical range, we also created maps with combined uv data sets from all array configurations observed at 270 and 341 GHz. This yields resolutions of 2.5"×2.1" and 1.45"×1.39" using natural and close to robust 0 baseline weightings at 270 GHz, and of 0.47"×0.45" and 0.39"×0.33" using natural and robust 0 baseline weightings at 341 GHz. The continuum sensitivity in these maps is 0.59 and 1.1 mJy beam$^{-1}$ at 270 and 341 GHz, respectively (natural weighting; see Figure S2 for continuum maps). We also searched the lower sideband (LSB) of the 341 GHz data for [NII] $^3P_2$−$^3P_1$ line emission. We find a possible small positive excess in flux density, but do not detect the line. We also do not detect HF $J$=2−1 in absorption (Figure S1).

## 2.7 IRAM 30m Telescope Goddard-IRAM Superconducting 2-Millimeter Observer (GISMO)

We used the GISMO 128-element 2 mm bolometer camera at the IRAM 30m telescope to observe a ~3'×4.5' field around HFLS3 at ~18" resolution (PI: Perez-Fournon). The source was observed for six 10 min scans on 2012 April 18 and 22, yielding an effective on source integration time of ~1 hr. The radio quasars 1637+574 and 1641+399 were observed before and after the source to derive pointing corrections for the telescope. The resulting astrometric error is below 3" r.m.s. Absolute fluxes were derived relative to Neptune (6.20 Jy on 2012 April 11), and are estimated to be accurate to 5%. The data were reduced and calibrated using version 2.12-2 of the CRUSH package.[44,45] We detect HFLS3 at 7.9σ significance at a central frequency of 150 GHz, yielding a flux density of 2.93±0.37 mJy (Figure S8).

## 2.8 William Herschel Telescope (WHT) and Gran Telescopio Canarias (GTC)

We have observed HFLS3 in white light and in the SDSS $g$, $r$, $i$, and RGO $Z$ filters (central wavelengths of 484.4, 622.8, 779.6, and 874.8 nm), using the ACAM instrument at the 4.2m WHT (8.3' diameter field of view, 0.25" size pixels) on 2010 August 7, with total exposure times of 20, 45, 40, 60, and 15 minutes, respectively, and with 0.9" seeing (PI: Perez-Fournon). Further broad-band imaging was obtained in the near-infrared $K_s$ filter (2.2 μm), using the LIRIS instrument (4.27'×4.27' field of view, 0.25" size pixels), on 2011 April 21. The total integration time was 59 minutes, and the seeing was 0.65". The reduction of the ACAM images was carried out using standard procedures in IRAF. The reduction of the LIRIS images was carried out using the IAC's IRAF lirisdr task.[a] Optical images were smoothed with at 1.1" Gaussian. The $K_s$ image was smoothed with a Gaussian filter of σ=0.25" (1 pixel) width.

Further, deeper optical imaging observations were carried out in the SDSS *g*, *r*, *i*, and *z* bands using the Optical System for Imaging and low Resolution Integrated Spectroscopy (OSIRIS) instrument on the GTC 10.4m telescope (7.8'×7.8' field of view at 0.365-1.05 μm, 0.25" size pixels; PI: Perez-Fournon).[b] Observations were carried out between 2011 June 29 and August 3, with total exposure times of 45, 45, 360, and 360 minutes, with 1.1", 0.96", 0.85", and 0.98" seeing in the final stacked images in the *g*, *r*, *i* and *z*-bands, respectively. Each observing block consisted of many dithered short exposures with a total integration time of 45 minutes. The reduction of the OSIRIS images was carried out using IRAF.

The astrometric calibration of the WHT and GTC images was carried out using the Graphical Astronomy and Image Analysis Tool (GAIA) included in the Starlink astronomical software package.[c,d] The LIRIS $K_s$ image was tied to 2MASS stars, and is estimated to be accurate to ~0.2" (r.m.s. of the fit). To achieve the best relative astrometric accuracy, we have carefully registered all optical images (and *Spitzer*/IRAC) to the reference frame of the $K_s$ LIRIS image. This was achieved by extracting a catalogue of bright sources within a 1.2' radius region around HFLS3 from the LIRIS $K_s$ image using SExtractor.[46] This source catalog was then used in GAIA to calculate an astrometric solution for the WHT ACAM, GTC OSIRIS, and *Spitzer*/IRAC images. The r.m.s. residuals of the astrometric fits are 0.16"–0.22" in all bands.

The photometric calibration of the optical images was carried out based on the Sloan Digital Sky Survey (SDSS) DR8 Photometric Catalog.[47] 2MASS was used for the near-infrared photometric calibration.[48]

We detect significant emission ~0.65" north of the submillimeter position of HFLS3 in all optical bands (Figure S9). This emission is not from HFLS3, but originates from a compact, faint *z*=2.091 galaxy, dubbed G1B (see below for redshift determination). We detect extended emission close to the position of HFLS3 in the $K_s$ band at 0.65" resolution, which originates from both G1B and HFLS3. No emission is detected at the position of HFLS3 in the optical bands, consistent with total absorption due to a Gunn-Peterson trough at observed-frame <892 nm, and significant dust obscuration longward of the Lyman break. Based on GALFIT models[49] we find that G1B can be fitted well with a galaxy profile in the optical bands without significant residuals. We used the same GALFIT models for optical photometry of G1B. SExtractor was used for $K_s$ band photometry to determine the combined near-infrared flux of both sources.

### 2.9 Keck Second Generation Near-Infrared Camera (NIRC2)

We obtained a 6.1 hr (276×80 s) $K_S$-band image of HFLS3 on 2011 April 13 and 2012 June 5, using the NIRC2 camera (0.04" size pixels, 40" field-of-view) and the Keck II laser guide-star adaptive-optics system (PIs: Fu, Riechers).[50] An $R$ = 14.7 magnitude star 43" north of HFLS3 served as the tip-tilt reference star. The estimated Strehl ratio at the source position is ~26%. The outside seeing was ~0.4" (2011) and ~0.8" (2012) in the optical. The adaptive optics corrections yielded point spread functions with effective FWHM of typically 2.2 pixels (2011) and 3.0 pixels (2012), as measured for point sources in the observed field. We thus have smoothed the co-added image with a Gaussian of 0.1" FWHM (Figures 3 and S10).

We clearly resolve the blend of HFLS3 and the *z*=2.1 galaxy G1B to the north detected in the seeing-limited WHT LIRIS image. G1B appears compact and barely resolved in this image, suggesting that it is a relatively small, low-mass galaxy (consistent with its compactness and brightness in the optical images). HFLS3 is detected, and spatially resolved. Using SExtractor aperture photometry, we estimate that HFLS3 contributes ~37%±4% to the combined $K_s$ band flux of both sources. Using the absolute flux scale from the seeing-limited WHT $K_s$ band image, this translates to a $K_s$ band flux density of 1.823±0.305 μJy, or 23.25±0.17 mag (AB) for HFLS3 (errors of all calibration steps were added in quadrature).

[a] http://www.iac.es/galeria/jap/lirisdr/LIRIS_DATA_REDUCTION.html

[b] http://www.gtc.iac.es/en/pages/instrumentation/osiris.php

[c] http://astro.dur.ac.uk/~pdraper/gaia/gaia.html

[d] http://starlink.jach.hawaii.edu/starlink

### 2.10 Keck Low-Resolution Imaging Spectrometer (LRIS)

HFLS3 was observed in two separate observing runs using the Keck I 10-m telescope Low-Resolution Imaging Spectrograph (LRIS; PI: Bridge).[51] For these data we used a 1.5" slit, the D560 dichroic with the 600/4000 grism (blue side) and the 400/8500 grating (red side). Binning the data 1×2 (spatial pixel×spectral pixel) on the blue side results in dispersions of 1.26 and 1.16 angstroms per pixel, respectively. The data were pre-processed and wavelength and flux calibrated using standard IRAF procedures. First observations were taken under seeing conditions of 0.8" on 2010 July 14, with a slit position centered on the SMA 1.1 mm coordinates. A total integration time of 20 minutes revealed weak CIV and tentative OIII] emission lines at a redshift of $z$=2.091, on top of weak continuum emission at position B in the slit (Figure S11). The position and brightness of the line and continuum emission is consistent with the galaxy G1B 0.65” to the north of HFLS3. In addition, Ly-$\alpha$ emission at $z$=2.202 is detected from a nearby galaxy at position A in the slit, 2.8” away from G1B (two additional bright sources on the slit, C and D, were used for positional referencing only). Additional spectra were taken on 2011 May 28 to confirm these lines, however the seeing was 1.0”−1.8" throughout the additional 150 min integration time, and therefore did not improve the significance of the line detections.

At $z$=6.3369, the Ly-$\alpha$ line is redshifted to 891.9 nm, which is within the range of our spectrum. However, we do not see any evidence for line or continuum emission from HFLS3 in our LRIS spectra. These findings are consistent with our broad-band photometry, and suggest that both Lyman-$\alpha$ and the continuum are extinguished due to high dust obscuration.

### 2.11 Wide-Field Infrared Survey Explorer (WISE)

The position of HFLS3 was covered by the Preliminary Release Catalog observations of WISE, which contains 57% of the sky.[52] The source remains undetected in the 3.4, 4.6, 12, and 22 μm bands (W1-W4) down to 5σ (r.m.s.) sensitivity levels of 0.08, 0.11, 0.8 and 4 mJy, respectively.

### 2.12 *Spitzer Space Telescope* InfraRed Array Camera (IRAC)

HFLS3 was observed for 1hr on source each in the 3.6 and 4.5μm bands (IRAC channel 1 and 2) on 2012 March 21 (warm mission; PI: Viera). Data reduction was performed using the MOPEX package and standard procedures. Photometry was extracted with the SExtractor package. Given the resolution of *Spitzer*, we detect a partial blend of the z=2.1 galaxy G1B and HFLS3 (Figure S9). After matching the reference frames, we used the positions of G1B in the optical and HFLS3 at longer wavelengths as priors for our de-blending procedures based on GALFIT models of both sources. The r.m.s. residuals of the astrometry matching are typically <0.2” between individual bands including IRAC, which is much smaller than the spatial separation of G1B and HFLS3 (see above). Despite the fact that the emission in the IRAC bands is dominated by G1B, it thus is possible to subtract off its contribution and measure the fluxes of HFLS3.

## 3. Determination of Physical Properties

### 3.1 SED Fitting

#### 3.1.1 Modified Black-Body Fitting

To determine the SED properties of HFLS3, we have fitted the continuum photometry data in Table S1 with the redshift fixed to $z$=6.3369. We fit modified black-body (MBB) functions to the continuum data between observed-frame 250 μm and 3 mm, which are expected to be dominated by thermal dust emission. In this framework, the flux density $S_\nu$ of the thermal emission at frequency ν scales with the Planck function $B_\nu$ and a frequency-dependent optical depth factor as $S_\nu \propto (1-\exp(-\tau_\nu))\, B_\nu(T)$, where $\tau_\nu=(\nu/\nu_0)^\beta$, and $\nu_0$ is the frequency where the optical depth reaches unity.[53] We join this functional form to a $\nu^{-\alpha}$ power law on the blue side of the SED peak with slope α. The fit parameters thus are the dust temperature $T_d$, the power law slope of the extinction curve $\beta$, the wavelength $\lambda_0=c/\nu_0$ where the optical depth reaches unity, and an overall normalization, which we choose to be the observed-frame 500 μm flux density $S_{500\mu m}$. The choice of $S_{500\mu m}$ instead of multiplying by an overall multiplication factor is motivated by considerably better numerical stability. From this fit, we can derive an overall (far-)infrared luminosity $L_{\rm (F)IR}$ and dust mass $M_{\rm d}$. We use an affine-invariant Markov Chain Monte Carlo (MCMC) sampler (*emcee*)[54] to compute 500,000 steps from 250 samplers. Our approach fully marginalizes over the optical depth, and takes covariances between the input photometry into account. The autocorrelation length is <50 steps for all parameters, i.e., the chains converge very well. The resulting constraints are shown in Figure S12.

Our best fit has a $\chi^2$ of 31.5 for 23 degrees of freedom. We find $\beta$=1.92±0.12 and $T_{\rm d}$=55.9$^{+9.3}_{-12.0}$ K at a best-fit 500 μm flux density of 44.0$^{+4.7}_{-4.6}$ mJy. The power-law slope to fit the blue side of the SED is only poorly constrained by the data ($\alpha$>2.5 at 95.4% probability). Our calculations also provide an upper limit for $\lambda_0$=c/$\nu_0$<1,194 μm (162.7 μm rest-frame) at 95.4% confidence level, with a best-fit value of 733$^{+326}_{-467}$ μm (99.9$^{+44.4}_{-63.6}$ μm rest-frame). This suggests that the observed emission lines at $\nu_{\rm obs}$>1.2 mm do not require extinction corrections in line excitation models. The fact that we can only place an upper limit on $\lambda_0$ has little effect on the estimated $L_{\rm (F)IR}$ and $M_{\rm d}$. Our fits imply an infrared (IR; 8−1000 μm) luminosity of $L_{\rm IR}$=(4.16$^{+0.27}_{-0.29}$)×10$^{13}$ L$_{\rm sun}$, and a far-infrared (FIR; 42.5−122.5 μm) luminosity of $L_{\rm FIR}$=(2.86$^{+0.32}_{-0.31}$)×10$^{13}$ L$_{\rm sun}$. We estimate the dust mass using:

$$M_{\rm d}=S_\nu\, D_{\rm L}^2\, [(1+z)\, \kappa_\nu\, B_\nu(T)]^{-1}\, \tau_\nu\, [1-\exp(-\tau_\nu)]^{-1},$$

assuming a mass absorption coefficient of $\kappa_\nu$=2.64 m$^2$ kg$^{-1}$ at 125 μm.[55] This yields a dust mass of $M_{\rm d}$=(1.31$^{+0.32}_{-0.30}$)×10$^9$ M$_{\rm sun}$. The error bars do not include uncertainties in $\kappa_\nu$, which are at least a factor of 3.

### 3.1.2 Template Fitting

Besides direct MBB fitting, we also used $\chi^2$ minimization routines to fit galaxy templates to the photometry of HFLS3. We selected templates for the nearby starburst galaxy M82, the nearby ultra-luminous infrared galaxy Arp 220, the *z*=1.44 red galaxy HR10, and the gravitationally lensed *z*=2.32 galaxy SMM J2135−0102 (the Cosmic Eyelash; Figure 2).[56,57,58] The FIR SED shape is not consistent with HR10 or the Eyelash. The near/mid-infrared to FIR flux ratio is only marginally consistent with M82. However, the overall SED shape is fairly consistent with Arp 220. The differences in the FIR SED shape relative to Arp 220 can be explained with a lower dust optical depth in HFLS3. The dust in Arp 220 gets optically thick at wavelengths shortward of rest-frame 234±3 μm,[20] whereas HFLS3 is optically thin longward of rest-frame 162.7 μm (95.4% confidence level; see Sect. 3.1.1). We also attempted to fit the SED with template libraries,[59,60] but this did not yield any significant additional constraints.

### 3.1.3 Optical Depth Fitting for other High Redshift Sources

Using the same calculations as above, we fitted the optical depth of other high-redshift galaxies. For the Cosmic Eyelash, our fit yields a $\chi^2$ of 13.6 for 7 degrees of freedom. We measure $\lambda_0$=212±24 μm in the rest frame. For the z=4.29 galaxy HFLS1 discovered as part of our survey, we obtain a $\chi^2$ of 11.2 for 5 degrees of freedom, and $\lambda_0$=200$^{+29}_{-25}$ μm in the rest frame. For the z=4.44 galaxy HFLS5, we obtain a $\chi^2$ of 3.6 for 5 degrees of freedom, and $\lambda_0$=198±52 μm in the rest frame. For the z=2.96 galaxy HLSW-01,[61] we obtain a $\chi^2$ of 6.8 for 6 degrees of freedom, and $\lambda_0$=197±19 μm in the rest frame. Thus, $\lambda_0$~200 μm is found for these high-redshift galaxies, with individual values that are compatible with that measured in Arp 220. Taken at face value, this would indicate that the optical depth in HFLS3 is somewhat lower than typical for ULIRGs and high-z massive starburst galaxies.

## 3.2 Star Formation Rate

We adopt a (top-heavy) Chabrier stellar initial mass function (IMF),[62] to derive the star formation rate of HFLS3 from the FIR luminosity based on a Kennicutt conversion:[63]

$$\mathrm{SFR[M_{sun}yr^{-1}]} = 1.0\times10^{-10}\, L_{\rm FIR}\, [\mathrm{L_{sun}}]\,.$$

For $L_{\rm FIR}$=(2.86$^{+0.32}_{-0.31}$)×10$^{13}$ L$_{\rm sun}$ as derived from our SED fits for HFLS3, this suggests a star formation rate (SFR) of ~2,900 M$_{\rm sun}$yr$^{-1}$. This SFR would be by a factor of ~1.7 higher when adopting a standard Salpeter IMF (i.e., ~4,900 M$_{\rm sun}$yr$^{-1}$),[63] and by a factor of ~2.6 lower when adopting a flat Baugh IMF (i.e., ~1,100 M$_{\rm sun}$yr$^{-1}$).[64,65] A flat IMF may be preferred to explain the formation of z~2 submillimeter galaxies.[64] On the other hand, present-day elliptical galaxies show evidence for bottom-heavy IMFs.[66,67,68] Assuming that submillimeter galaxies are the ancestors of these galaxies, it may be expected that their IMF could also be bottom-heavy. This, however, could imply an even higher star formation rate than in any of the cases quoted above. Assuming, that the theoretical limit of $\Sigma_{SFR}$ = 1,000 $M_{sun}yr^{-1}\,kpc^{-2}$ applies to HFLS3, IMFs that are more bottom-heavy than Salpeter are inconsistent with our observations. We adopt a Chabrier IMF for the remainder of this work.

## 3.3 Gas Mass

### 3.3.1 Molecular Gas Mass

To derive a total molecular gas mass from the CO luminosity of HFLS3, we assume a low conversion factor of $\alpha_{CO} = M_{gas}/L'_{CO}$ = 1 $M_{sun}$ (K kms$^{-1}$ pc$^2$)$^{-1}$ from CO line luminosity to $M_{gas}$, as commonly used for ULIRGs (typical values of $\alpha_{CO}$=0.6–1.0, with outliers out to 0.3–1.3 $M_{sun}$ (K kms$^{-1}$ pc$^2$)$^{-1}$)[27] and submillimeter-selected galaxies at high redshift.[69] The CO luminosity of HFLS3 is $L'_{CO}$=(1.04±0.09)×10$^{11}$ K kms$^{-1}$ pc$^2$. This suggests a total molecular gas mass of $M_{gas}$=(1.04±0.09)×10$^{11}$$M_{sun}$. If we were to adopt a high conversion factor of $\alpha_{CO}$ = 4 $M_{sun}$ (K kms$^{-1}$ pc$^2$)$^{-1}$ as found for nearby spiral galaxies including the Milky Way and high-redshift disk galaxies,[70] the molecular gas mass would exceed the dynamical mass. Any $\alpha_{CO}$ > 2.7 $M_{sun}$ (K kms$^{-1}$ pc$^2$)$^{-1}$ (>2.0 $M_{sun}$ (K kms$^{-1}$ pc$^2$)$^{-1}$ when including $M_d$, $M_*$, $M_{CI}$, and $M_{HI}$) are thus disfavored by our observations.

### 3.3.2 Neutral Atomic Carbon Mass

To derive the neutral atomic carbon mass in HFLS3 (in $M_{sun}$), we use the following equation:

$$M_{CI} = 5.706\times10^{-4}\, Q(T_{ex})\, 1/3\, \exp(23.6/T_{ex})\, L'_{[CI]}\,,$$

where $Q(T_{ex}) = 1 + 3\exp(-T_1/T_{ex}) + 5\exp(-T_2/T_{ex})$ is the [CI] partition function, $T_1$=23.6 K and $T_2$ = 62.5 K are the energies above the ground state for both [CI] lines, and $L'_{[CI]}$ is the line luminosity in the lower [CI] line.[71] The [CI] luminosity of HFLS3 is $L_{[CI]}$=3.0±1.9×10$^8$ $L_{sun}$. Assuming a typical [CI] excitation temperature of 30 K and a brightness temperature ratio of 0.5 between the [CI] levels, we find an atomic carbon mass of 4.5×10$^7$ $M_{sun}$.

### 3.3.3 Atomic Gas Mass

To determine the atomic gas (HI) mass in HFLS3 (in $M_{sun}$) based on [CII], we use the following equation (which assumes that the [CII] emission is optically thin):

$$M_{HI} = 0.77\,(0.7\,L_{[CII]})\,(1.4 \times 10^{-4}/X_{C+})$$
$$\times(1{+}2\exp(-91\text{ K}/T){+}n_{crit}/n)\,(2\exp(-91\text{ K}/T))^{-1}\,,$$

where $X_{C+}$=1.4 x 10$^{-4}$ is the [CII] abundance per hydrogen atom, and $n_{crit}$=2.7×10$^3$ cm$^{-3}$ is the critical density of [CII].[72] Assuming $n$=10$^{3.8}$ cm$^{-3}$ and $T$=144 K as measured for the density and kinetic temperature of the molecular gas (see Supplementary Section 4), we find $M_{HI}$=2.0×10$^{10}$ $M_{sun}$. This corresponds to ~20% of the molecular gas mass.

## 3.4 Stellar Mass

From fitting population synthesis models[73] to the optical to mid-IR data of the foreground galaxy G1B using *hyperz*,[74] we obtain a stellar mass of $M_*$=4.5$^{+1.1}_{-4.2}$×10$^{10}$ $M_{sun}$. We also have fitted population synthesis models to HFLS3, but given the high redshift and high dust obscuration of the source, the data allow for a broad range in $M_*$ between 2×10$^8$ and 6×10$^{11}$ $M_{sun}$, with an interquartile range of 0.2–2×10$^{10}$ $M_{sun}$. Given the mass of the dust and different gas phase masses, stellar masses of >1.5×10$^{11}$ $M_{sun}$ would exceed the dynamical mass (even when assuming that no dark matter is present; under the assumption of a dark matter fraction of 25%, this limit would decrease to >7.7×10$^{10}$ $M_{sun}$), and thus, are disfavored due to other observational constraints. Likewise, given the considerable dust mass, stellar masses at the low end of the formally allowed range are disfavored, assuming that a significant fraction of the dust was produced by stars in the galaxy. Making the conservative assumption that the galaxy contains at least as much mass in stars as in dust disfavors any stellar masses of <1.3×10$^9$ $M_{sun}$ (which is close to the lower limit of the interquartile range). The interquartile range would suggest that the baryonic mass in HFLS3 is dominated by molecular gas rather than stars, and thus, that the ongoing starburst is in a relatively early phase.

We have also fitted the SED of HFLS3 with *CIGALE*,[75] using dust emission templates (i.e., taking the full spectral energy distribution into account),[59]

assuming a metallicity of 0.5 $Z_{sun}$, and three different star formation histories (constant, and exponentially declining/rising). The results for the stellar mass only weakly depend on the assumed star formation history, suggesting $M_*$=3.7×10$^{10}$ $M_{sun}$ at an $A_V$ of 3.29. The $M_*$ estimate for a Salpeter IMF is by a factor of ~1.8 higher (i.e., $M_*$=6.8×10$^{10}$ $M_{sun}$) than for a Chabrier IMF (which is given here). Differences for more extremely top- and bottom-heavy IMFs are typically ~1 dex.[76] For Lyman-break galaxies at similar redshift, the combined uncertainty due to metallicity and the assumed star formation history for determining the stellar mass is ~0.3 dex. The uncertainty due to a possible contribution of nebular emission lines (which *CIGALE* takes into account) to the broad-band fluxes for the same galaxies is ~0.2−0.4 dex.[77] Given the high level of dust obscuration in HFLS3, these thus are unlikely to be dominant sources of uncertainty. Both independent estimates of $M_*$ thus are consistent within the (considerable) relative uncertainties. We adopt the value derived from the *CIGALE* fit in the following. This value is about half of the median stellar mass of z~2 submillimeter galaxies of ~7×10$^{10}$ $M_{sun}$.[78]

### 3.5 Dynamical Mass

To determine the dynamical mass of the system (in $M_{sun}$), we use the "isotropic virial estimator":

$$M_{\rm dyn} = 2.8\times10^5\,(\Delta v_{\rm FWHM})^2\,r_{1/2}\,,$$

where $\Delta v_{\rm FWHM}$ is the CO line width in kms$^{-1}$ (we here adopt the width of the CO $J$=6−5 line), and $r_{1/2}$ is the half light radius in kpc (we here adopt the major axis FWHM radius of the [CII] emission, which is consistent with estimates from the lower-resolution CO observations).[32] In this equation, a scaling factor appropriate for submillimeter galaxies (which are typically disturbed systems) is adopted. The scaling factor for a rotating disk at an average inclination would be ~1.5 times smaller, but this estimator was found to agree well with masses derived from detailed Jeans modeling for massive high redshift galaxies.[32] With these assumptions, we find a dynamical mass of 2.7×10$^{11}$ $M_{sun}$. Adopting the $M_{gas}$, $M_d$, $M_*$, $M_{CI}$, and $M_{HI}$ as determined above, this suggests a dark matter fraction of $f_{DM}$~40% of the dynamical mass, which would be higher than the typically adopted 20%−25% for z~1−2 disk galaxies and z~2 SMGs.[70,15] Given the remaining uncertainties in $M_*$ and other quantities entering this calculation, the significance of this finding is only moderate.

### 3.6 Gas Mass Fraction

We define the gas mass fraction as $f_{\rm gas} = M_{\rm gas}/M_{\rm dyn}$. For HFLS3, we find $f_{\rm gas}$ ~ 40%. This is comparable to what is found in submillimeter-selected starbursts and massive star-forming galaxies at z~2,[15,16] but ~3 times higher than in nearby ULIRGs like Arp 220, and >30 times higher than in the Milky Way. In particular, this value is consistent with, but towards the high end of the plateau in $f_{\rm gas}$ observed between 2<z<5 (when adjusted to the same definition),[79] and extends it toward z>6. This suggests that, the gas supply rates in the most massive halos at z>6 are high enough to support rapid stellar mass buildup, as witnessed in its peak phase in HFLS3 (as shown by the high SFR and $\Sigma_{SFR}$, which imply a close-to-maximal gas consumption rate). It also implies that very high redshift massive starburst galaxies like HFLS3 are not yet very evolved in terms of their conversion of gas into stars, showing that a major fraction of the stellar mass has yet to grow.

### 3.7 Radio-FIR Correlation

We calculate the monochromatic rest-frame 1.4 GHz radio luminosity assuming:

$$L_{\rm 1.4GHz} = 4\pi D_{\rm L}^2\,(1+z)^{-(1+\alpha_{\rm 1.4GHz})}\,S_{\rm 1.4GHz}\,,$$

where $\alpha_{1.4GHz}$ = -0.75 is used as the radio spectral index,[80,81] and $D_L$ is the luminosity distance. This gives $L_{1.4GHz}$ = 1.7±0.3×10$^{25}$ W Hz$^{-1}$. The radio-FIR correlation can be expressed via the so-called $q$-parameter:[82]

$$q = \log_{10}(L_{\rm FIR}/9.8 \times 10^{-15}\,{\rm L_{sun}}) - \log_{10}(L_{\rm 1.4GHz}/{\rm W\,Hz^{-1}})\,.$$

For HFLS3, this yields $q$=2.33±0.43, which agrees very well with the radio-FIR correlation for nearby star-forming galaxies ($q$=2.3±0.1),[83] and with that for SMGs and HerMES sources with firm radio identification in general ($q$=2.4±0.1).[84,81] The radio luminosity of HFLS3 thus is fully consistent with non-thermal synchrotron emission from supernova remnants (SNR) within the star-forming regions, rather than requiring the presence of a radio-loud active galactic nucleus.

## 4. Excitation modeling

### 4.1 CO Excitation

We have modeled the CO excitation ladder in HFLS3 using the RADEX radiative transfer code, assuming an escape probability formalism for spherical shells (i.e., large velocity gradient modeling).[85] As input parameters, we use the average CO line width as determined from the CO $J$=2−1 and 6−5 lines of 659 $kms^{-1}$, and a [CO/$H_2$] abundance ratio of $10^{-4}$. We further fixed the cosmic microwave background temperature at $z$=6.34 to $T_{CMB,z}$=20.02 K. We compute CO intensities for a large model grid in kinetic temperature ($T_{kin}=10^{0.7}–10^{3.7}$ K), molecular hydrogen density density ($n(H_2)=10^{2.0}–10^{6.0}$ $cm^{-3}$), ratio of CO column density and velocity width ($N_{CO}/dv=10^{15}–10^{20}$ $cm^{-2}$ $(kms^{-1})^{-1}$), and CO equivalent disk filling factor ($\Phi_A=10^{-4.0}–10^{-1.0}$). From the resulting model grid, we generated likelihood distributions for all model parameters. As priors, we assume that the predicted molecular gas mass cannot exceed the dynamical mass of $M_{dyn}=2.7\times10^{11}$ $M_{sun}$, and that the CO column length is smaller than or equal to the observed diameter of the [CII]- and CO-emitting region. The former prior sets an effective upper limit on $N_{CO}$. The latter prior sets an effective lower limit on $n(H_2)$ and an upper limit on $N_{CO}$. Calibration errors for individual line flux measurements are added in quadrature. Also, given the limited observational constraints, we assume that all observed CO lines can be described by a single excitation component. To obtain likelihood distributions for a single parameter, we integrate the likelihood matrix over all other dimensions. For comparison, we also calculate the maxima in the likelihood distribution of each parameter ("1D Max"), and its maximum in the best-fit solution for $T_{kin}$, $n(H_2)$, $N_{CO}$, and $\Phi_A$ combined ("4D Max").[86] We define the thermal gas pressure as $P=P_{Th}/k_B=T_{kin}\times n(H_2)$, and the beam-averaged column density as $<N_{CO}> = N_{CO}\times\Phi_A$. We also derive likelihood distributions for the velocity gradient $dv/dr$. The model-predicted median parameters and corresponding 1σ ranges are summarized in Table S3. The resulting maximum likelihood model is shown in Figure S13. One- and two-dimensional likelihood contours are shown in Figure S14. To investigate the presence of multiple gas components, we repeated the model calculations assuming two excitation components. These models did not provide a better fit to the observed CO line ladder than the single-component models within the uncertainties. For reference, the best-fit two-component model gives $T_{kin}$~100 and 250 K and $n(H_2)\sim10^{2.6}$ and $10^{5.0}$ $cm^{-3}$, respectively.

### 4.2 $H_2O$ Excitation

Given the limited constraints on the $H_2O$ excitation ladder (Figure S15), and the fact that the $H_2O$ line ratios in HFLS3 (within the errors) are consistent with those in Arp 220, we first explored a wider parameter space based on the $H_2O$ excitation ladder of Arp 220, and used our findings as input to models of HFLS3. We used RADEX to produce $H_2O$ radiative transfer models. The parameters and ranges explored by our models are $T_{kin}=10^{1.3}–10^{3.3}$ K, $n(H_2)=10^{3.0}–10^{11.0}$ $cm^{-3}$, $N_{H2O}=10^{16}–10^{25}$ $cm^{-2}$, $\Phi_A=10^{-5}–1$, and [$H_2O/H_2$]=$10^{-9}–10^{-5}$. As priors, we assume that the predicted molecular gas mass cannot exceed the dynamical mass, and that the $H_2O$ column length is smaller than or equal to the observed diameter of the CO and $H_2O$-emitting region. We also fixed the $H_2O$ ortho-to-para ratio to 3, and reject RADEX runs where the optical depth of all lines is outside the range $\tau=10^{-10}–10^2$. We generated posterior likelihood distributions for all model parameters using the nested sampling routine MULTINEST, which we used to call RADEX and create models "on-the-fly".[87] MULTINEST is a Bayesian tool similar to MCMC, but is more efficient when dealing with multi-modal regions in parameter space. It also calculates values for the Bayesian evidence. For both Arp 220 and HFLS3, [$H_2O/H_2$], and thus, $M_{gas}$, are poorly constrained by the $H_2O$ data alone. Solutions with $T_{kin}$~100−150 K are preferred, comparable to the best-fitting values for CO in HFLS3. However, due to the high critical densities of the $H_2O$ lines, gas densities of $>10^{8.5}$ $cm^{-3}$ are preferred (black contours in Figure S16). Such high gas densities over extended regions are unphysical. Also, the maximum likelihood solutions only poorly reproduce the observed line ratios in Arp 220 and HFLS3, in particular between directly connected energy levels (as higher energy levels are observed to have higher line fluxes in both Arp 220 and HFLS3; Figure S15). Our models thus suggest that collisional excitation is unlikely to be the main mechanism to produce the observed $H_2O$ excitation in Arp 220 and HFLS3.

To further explore whether or not collisional excitation is important for lower energy levels, we carried out a separate modeling run, taking only fluxes and limits for lines with $E_{upper}/k_B$ <~200 K

into account. Solutions with less extreme gas densities of $10^7$–$10^8$ cm$^{-3}$ are preferred (red contours in Figure S16), but at significantly higher $T_{\rm kin}$ of few hundreds to >1000 K. The best-fitting models still only poorly reproduce the observed line ratios of directly connected energy levels. Thus, it appears unlikely that collisional excitation dominates the lower-energy $H_2O$ levels either.

We detect emission from $H_2O$ lines with upper level energies of $E/k_{\rm B}$ > 300–450 K, which would require gas densities in excess of $10^{8.5}$ cm$^{-3}$ to reproduce the observed relative line strengths through collisions. However, the $J_{\rm KaKc}$=$3_{21}$ and $4_{22}$ energy levels of ortho- and para-$H_2O$ can be efficiently populated by 75 and 58 μm infrared photons through absorption in the $J_{\rm KaKc}$=$2_{12}$–$3_{21}$ and $3_{13}$–$4_{22}$ ortho- and para-$H_2O$ transitions, respectively, which then can produce the observed emission line strength in the cascading transitions. The 75 μm transition coincides with the observed peak of the SED, allowing for a high pumping efficiency. The 58 μm transition falls on the Wien tail of the SED. In combination with the higher energy of the lower level of the pumping transition (and thus, lower population through collisions) compared to the ortho-$H_2O$ channel, this explains the much lower strength of high-level para-$H_2O$ lines. The relative strength of all detected $H_2O$ lines is consistent with those observed in Arp 220 (Fig. 3).[20] The 75 μm and 58 μm pumping transitions were detected in absorption in Arp 220, with the 75 μm feature being more than twice as deep.[21] This is also consistent with the fact that we detect the OH $^2\Pi_{1/2}(3/2-1/2)$ feature at 163 μm in emission, which has an upper level energy comparable to the $J_{\rm KaKc}$=$3_{21}$ level of ortho-$H_2O$, but can also be efficiently pumped by 53.3 and 35 μm photons, as observed in Arp 220.[21]

Our models thus suggest that collisional excitation is unlikely to be the main mechanism to produce the observed $H_2O$ excitation in HFLS3, which instead may be enhanced by the infrared radiation field in the star-forming regions. This is consistent with observations and models of very high energy levels of $H_2O$ in Arp 220, which are observed in absorption.[21]

## 5. Gravitational Lensing

To investigate the possibility of gravitational lensing, we have used the CO luminosity-line width relation for lensed submillimeter galaxies to estimate the gravitational lensing magnification factor,[88] i.e.,

$$\mu_{\rm L} = 3.5\times10^{-11}\, L'_{\rm CO}\times(\Delta v_{\rm FWHM}/400\ {\rm kms^{-1}})^{-1.7},$$

where $L'_{\rm CO}$ is given in units of K kms$^{-1}$ pc$^2$. Taking the average $\Delta v_{\rm FWHM}$ of the CO $J$=2–1 and 6–5 lines, this suggests $\mu_{\rm L}$=1.5±0.7, with the error for $L'_{\rm CO}$ and the median error for $\Delta v_{\rm FWHM}$ added in quadrature to the estimated 40% uncertainty of the $L'_{\rm CO}$–$\Delta v_{\rm FWHM}$ relation for the $\Delta v_{\rm FWHM}$ of HFLS3. This result is consistent with no strong gravitational lensing within the uncertainties, which is also consistent with the compact measured size of the submillimeter continuum emission.

From fitting population synthesis models to the optical/near-IR data of the foreground galaxy G1B, we obtain a stellar mass of $M_*$=$4.5^{+1.1}_{-4.2}\times10^{10}$ M$_{\rm sun}$. Based on standard relations, this suggests a relatively low velocity dispersion of 100±30 kms$^{-1}$. This translates to a small Einstein radius of $0.11''^{+0.05''}_{-0.04''}$. At the distance of HFLS3, this suggests a lensing magnification factor of $\mu_{\rm L}$=$1.24^{+0.14}_{-0.11}$ (16$^{\rm th}$ and 84$^{\rm th}$ percentiles). Also, we do not detect a lensed counter image of HFLS3 on the opposite side of G1B in our (sub)millimeter data. From the high-resolution 1.2 mm continuum data, we obtain an upper limit on the image-to-counter-image peak brightness ratio of 0.1 (4σ). These findings are consistent with no strong gravitational lensing within the uncertainties.

Furthermore, we have used models of HerMES 250, 350, and 500 μm number counts to estimate the number density of sources with $S_{500\mu m}$>30 mJy.[7] At $z$>6 ($z$>5), these models suggest source densities of 0.014 deg$^{-2}$ (0.1 deg$^{-2}$) at such brightness levels, 60% (64%) of which are expected to be starburst galaxies. Only 29% (28%) of these starbursts are expected to be gravitationally lensed. These estimates are consistent with the space densities of ultra-red sources like HFLS3 within the relative uncertainties. Moreover, this suggests that a $z$>6 starburst like HFLS3, once detected, is less likely to be lensed than unlensed. Together, the above estimates are consistent with HFLS3 not being strongly gravitationally lensed, despite its relatively high observed brightness.

## 6. Supplementary Tables

### Table S1: Multi-wavelength photometry of HFLS3.

| Wavelength [μm] | Frequency [GHz] | Flux Density [mJy] | Error [mJy] | Observatory |
|---|---|---|---|---|
| 0.4686 | 640,000 | <0.052e-3 | | GTC (g) |
| 0.6165 | 485,000 | <0.083e-3 | | GTC (r) |
| 0.7481 | 400,000 | <0.052e-3 | | GTC (i) |
| 0.8931 | 335,000 | <0.157e-3 | | GTC (z) |
| 2.2 | 135,000 | 1.823e-3 | 0.305e-3 | WHT+Keck (Ks) |
| 3.4 | 88,000 | <0.08 | | WISE |
| 3.6 | 83,000 | 2.39e-3 | 0.25e-3 | IRAC |
| 4.5 | 67,000 | 3.16e-3 | 0.52e-3 | IRAC |
| 4.6 | 65,000 | <0.11 | | WISE |
| 12 | 25,000 | <0.8 | | WISE |
| 22 | 13,600 | <6 | | WISE |
| 70 | 4,300 | <2.0 | | PACS |
| 110 | 2,700 | <2.2 | | PACS |
| 160 | 1,900 | <4.0 | | PACS |
| 250 | 1,200 | 12.0 | 2.3* | SPIRE |
| 350 | 850 | 32.4 | 2.3* | SPIRE |
| 500 | 600 | 47.3 | 2.8* | SPIRE |
| 880 | 341 | 33.0 | 2.4 | SMA |
| 1,055.0 | 284.161 | 20.57 | 0.45 | Z-spec |
| 1,110 | 270 | 21.3 | 1.1 | SMA |
| 1,151.6 | 260.329 | 17.10 | 0.37 | Z-spec |
| 1,157.5 | 259.106 | 17.12 | 0.81 | PdBI |
| 1,157.5 | 259.106 | 13.9 | 1.9 | CARMA |
| 1,190.1 | 251.906 | 13.51 | 1.22 | CARMA |
| 1,196.8 | 250.490 | 14.07 | 0.60 | PdBI |
| 1,247.2 | 240.364 | 15.05 | 0.19 | PdBI |
| 1,247.7 | 240.284 | 14.17 | 0.38 | Z-spec |
| 1,317.2 | 227.604 | 10.13 | 0.29 | PdBI |
| 1,349.4 | 222.167 | 11.76 | 0.40 | Z-spec |
| 1,448.2 | 207.013 | 9.78 | 0.45 | Z-spec |
| 1,469.4 | 204.027 | 9.79 | 0.29 | PdBI |
| 1,821.4 | 164.598 | 6.57 | 0.18 | PdBI |
| 1,900.3 | 157.757 | 4.59 | 0.39 | PdBI |
| 2,000 | 150 | 2.93 | 0.37 | GISMO |
| 2,014.7 | 148.800 | 3.35 | 0.12 | PdBI |
| 2,121.2 | 141.328 | 3.22 | 0.12 | PdBI |
| 2,226.5 | 134.650 | 2.38 | 0.11 | PdBI |
| 2,633.9 | 113.819 | 1.25 | 0.09 | PdBI |
| 2,722.2 | 110.128 | 1.21 | 0.10 | PdBI |
| 2,722.2 | 110.128 | 1.59 | 0.12 | CARMA |
| 2,924.8 | 102.500 | 0.705 | 0.134 | CARMA |
| 3,181.0 | 94.246 | 0.527 | 0.078 | CARMA |
| 6,360.8 | 47.1311 | 0.139 | 0.030 | JVLA (Q) |
| 9,540.9 | 31.4217 | 0.0469 | 0.0093 | JVLA (Ka) |
| 19,081.5 | 15.7112 | <0.015 | | JVLA (U) |
| 214,000 | 1.4 | 0.059 | 0.011 | JVLA (L) |

*error bars on SPIRE fluxes are obtained from fitting and do not account for confusion noise, which is at least ~6mJy in all SPIRE bands.

**Table S2: Spectral line parameters (second entries in columns 3−6 are 1σ uncertainties).**

| Transition | Rest Frequency | Peak flux density | Velocity FWHM | Line Intensity | Line Luminosity | Obs. |
|---|---|---|---|---|---|---|
| | [GHz] | [mJy] | [km/s] | [Jy km/s] | [$10^{10}$ $L_l^{\#}$] | |
| **CO *J*=1-0** | 115.2712018 | 0.249 0.077 | 280 118 | 0.074 0.024 | 9.7 3.2 | JVLA* |
| **CO *J*=2-1** | 230.5380000 | 0.525 0.047 | 567 66 | 0.315 0.028 | 10.4 0.9 | JVLA |
| **CO *J*=3-2** | 345.7959899 | 0.692 0.090 | 977 160 | 0.717 0.094 | 10.5 1.4 | JVLA |
| **CO *J*=5-4** | 576.2679305 | | | <1.85 | <9.7 | CARMA |
| **CO *J*=6-5** | 691.4730763 | 3.44 0.86 | 752 232 | 2.74 0.68 | 10.0 2.5 | CARMA |
| **CO *J*=7-6** | 806.6518060 | 3.16 0.75 | 866 298 | 2.90 0.77 | 7.8 2.0 | CARMA |
| | | 2.14 0.21 | 979 145 | 2.22 0.25 | 6.0 0.7 | PdBI |
| **CO *J*=9-8** | 1,036.912393 | 3.48 0.57 | 497 107 | 2.77 0.45 | 4.5 0.7 | PdBI |
| **CO *J*=10-9** | 1,151.985452 | 3.94 2.01 | 937 165 | 3.91 1.59 | 5.2 2.1 | PdBI |
| **CO *J*=12-11** | 1,381.995105 | | | <18.1 | <16.6 | Z-spec |
| **CO *J*=13-12** | 1,496.922909 | | | <1.99 | <1.6 | PdBI |
| **CO *J*=14-13** | 1,611.793518 | | | <6.0 | <4.0 | Z-spec |
| **CO *J*=15-14** | 1,726.602507 | | | <8.0 | <4.7 | Z-spec |
| **CO *J*=16-15** | 1,841.345506 | | | <1.80 | <0.9 | PdBI |
| **CO *J*=17-16** | 1,956.018139 | | | <8.6 | <3.9 | Z-spec |
| **CO *J*=18-17** | 2,070.615993 | | | <10.5 | <4.3 | Z-spec |
| **CO *J*=19-18** | 2,185.134680 | | | <13.1 | <4.8 | Z-spec |
| **$H_2O$ $2_{11}$-$2_{02}$** | 752.033227 | 2.68 0.78 | 927 330 | 2.63 0.76 | 8.1 2.3 | CARMA |
| **$H_2O$ $2_{02}$-$1_{11}$** | 987.926764 | 2.56 0.31 | 805 129 | 2.19 0.33 | 3.9 0.5 | PdBI |
| **$H_2O$ $3_{12}$-$3_{03}$** | 1,097.364791 | 2.59 0.43 | 672 146 | 1.83 0.45 | 2.7 0.5 | PdBI |
| **$H_2O$ $3_{12}$-$2_{21}$** | 1,153.126822 | 2.51 2.08 | 937 165 | 2.49 1.74 | 3.3 2.3 | PdBI |
| **$H_2O$ $3_{21}$-$3_{12}$** | 1,162.911593 | 4.84 0.77 | 937 165 | 4.81 0.76 | 6.2 0.9 | PdBI |
| **$H_2O$ $4_{22}$-$4_{13}$** | 1,207.638714 | 1.56 0.60 | | 1.25 0.48 | 1.5 0.5 | PdBI* |
| **$H_2O$ $5_{23}$-$5_{14}$** | 1,410.618074 | | | <5.9 | <5.2 | Z-spec |
| **$H_2O$ $4_{13}$-$4_{04}$** | 1,602.219182 | | | 3.96 2.70 | 2.7 1.8 | Z-spec* |
| **$H_2O$ $2_{21}$-$2_{12}$** | 1,661.007637 | | | <1.92 | <1.2 | PdBI |
| **$H_2O$ $2_{12}$-$1_{01}$** | 1,669.904775 | | | <2.26 | <1.4 | PdBI |
| **$H_2O$ $3_{03}$-$2_{12}$** | 1,716.769633 | | | <6.7 | <4.0 | Z-spec |
| **$H_2O$ $6_{33}$-$6_{24}$** | 1,762.042791 | | | <6.2 | <3.5 | Z-spec |
| **$H_2O$ $6_{24}$-$6_{15}$** | 1,794.788953 | | | <7.5 | <4.1 | Z-spec |
| **$H_2O$ $7_{34}$-$7_{25}$** | 1,797.158762 | | | <7.5 | <4.1 | Z-spec |
| **$H_2O$ $5_{32}$-$5_{23}$** | 1,867.748594 | | | <6.8 | <3.4 | Z-spec |
| **$H_2O$ $5_{23}$-$4_{32}$** | 1,918.485324 | | | <7.3 | <3.5 | Z-spec |
| **$H_2O$ $3_{22}$-$3_{13}$** | 1,919.359531 | | | <7.5 | <3.6 | Z-spec |
| **$H_2O$ $4_{31}$-$4_{22}$** | 2,040.476810 | | | <10.5 | <4.4 | Z-spec |
| **$H_2O$ $4_{13}$-$3_{22}$** | 2,074.432305 | | | <11.6 | <4.7 | Z-spec |
| **$H_2O$ $3_{13}$-$2_{02}$** | 2,164.131980 | | | <11.4 | <4.2 | Z-spec |
| **$H_2O$ $3_{30}$-$3_{21}$** | 2,196.345756 | | | <16.0 | <5.8 | Z-spec |
| **$H_2O$ $5_{14}$-$5_{05}$** | 2,221.750500 | | | <16.8 | <5.9 | Z-spec |
| **$H_2O^+$ $2_{02}$-$1_{11}$** | | | | | | |
| **   *J*=5/2-3/2** | 742.0332 | | | <1.08 | <3.4 | CARMA |
| **   *J*=3/2-3/2** | 746.1938 | 1.64 0.99 | 542 410 | 0.94 0.57 | 3.0 1.8 | CARMA* |

| Transition | Rest Frequency | Peak flux density | Velocity FWHM | Line Intensity | Line Luminosity | Obs. |
|---|---|---|---|---|---|---|
| | [GHz] | [mJy] | [km/s] | [Jy km/s] | [$10^{10}$ $L_l^{\#}$] | |
| **OH $^2\Pi_{1/2}$ 3/2-1/2** | 1,834.74735/ 1,837.81682 | 7.82 0.80 | 1493 222 | 12.37 1.43 | 6.4 0.7 | PdBI |
| | | | | 5.98 3.25 | 3.1 1.7 | Z-spec* |
| **$OH^+$ $1_{1F}$-$0_{1F}$** | ~1,033.0582 | | | -0.56 0.18 | -0.92 0.29 | PdBI* |
| **$CH^+$ *J*=1-0** | 835.07895 | | | <1.1 | <2.8 | PdBI |
| **$CH^+$ *J*=2-1** | 1,669.15951 | | | <2.3 | <1.4 | PdBI |
| **$NH_3$ 3*K*a-2*K*s** | ~1,763.6496 | -1.56 0.51 | 648 290 | -1.07 0.37 | -0.60 0.21 | PdBI* |
| | | | | -5.68 2.83 | -3.2 1.6 | Z-spec* |
| **NH $2_2$-$1_1$** | ~1,958.2068 | | | -7.06 3.10 | -3.2 1.4 | Z-spec* |
| **HF *J*=2-1** | 2,463.42814 | | | <19.6 | <5.6 | SMA |
| **[CI] $^3P_2$-$^3P_1$** | 809.343500 | 1.82 1.30 | 276 240 | 0.53 0.37 | 1.4 1.0 | CARMA* |
| | | 0.76 0.45 | 419 341 | 0.34 0.21 | 0.91 0.57 | PdBI |
| **[CII] $^2P_{3/2}$-$^2P_{1/2}$** | 1,900.543 | 29.34 6.12 | 470 135 | 14.62 3.05 | 7.1 1.6 | CARMA |
| | | | | 9.99 2.86 | 4.8 1.4 | Z-Spec |
| | | | | 13.66 2.07 | 6.6 1.0 | PdBI |
| **[NII] $^3P_1$-$^3P_0$** | 1,461.132 | | | <6.3 | <5.2 | Z-spec |
| **[NII] $^3P_2$-$^3P_1$** | 2,459.380 | | | <19.6 | <5.7 | SMA |
| **[OI] $^3P_1$-$^3P_0$** | 2,060.068 | | | <7.4 | <3.0 | Z-spec |

#line luminosity is given in units of $L_l$ = K km/s pc$^2$

*tentative detections only, independent confirmation required

## Table S3: CO excitation modeling: parameters

| Parameter | Median value | 1σ range | 1D Max | 4D Max |
|---|---|---|---|---|
| **$T_{kin}$ [K]** | 143.59 | 113.64-202.21 | 138.04 | 138.04 |
| **$\log_{10}$ $n(H_2)$ [cm$^{-3}$]** | 3.84 | 3.63-4.08 | 3.80 | 3.80 |
| **$\log_{10}$ $N_{CO}$ [cm$^{-2}$]** | 20.84 | 20.63-21.03 | 20.92 | 21.02 |
| **$\log_{10}$ $\Phi_A$** | −1.78 | −1.88 - −1.68 | −1.70 | −1.70 |
| **$\log_{10}$ $P$ [K cm$^{-2}$]** | 6.02 | 5.82-6.24 | 6.12 | 6.12 |
| **$\log_{10}$ <$N_{CO}$> [cm$^{-2}$]** | 19.08 | 18.83-19.30 | 19.21 | 19.21 |
| **d$v$/d$r$ [kms$^{-1}$pc$^{-1}$]** | 1.4 | 0.8-3.5 | 1.3 | 1.3 |

**Table S4: Measured and derived source properties**

| Parameter | Value |
|---|---|
| $L'_{CO}$ | $1.04\pm0.09\times10^{11}$ K kms$^{-1}$pc$^{2}$ |
| $L_{CO}$ | $5.08\pm0.45\times10^{6}$ $L_{sun}$ |
| $L_{[CI]}$ | $3.0\pm1.9\times10^{8}$ $L_{sun}$ |
| $L_{[CII]}$ | $1.55\pm0.32\times10^{10}$ $L_{sun}$ |
| $L_{FIR}$ | $2.86^{+0.32}_{-0.31}\times10^{13}$ $L_{sun}$ |
| $M_{gas}$[a] | $1.0\times10^{11}M_{sun}$ |
| $M_{CI}$[b] | $4.5\times10^{7}$ $M_{sun}$ |
| $M_{HI}$[c] | $2.0\times10^{10}$ $M_{sun}$ |
| $M_{dust}$ | $1.31^{+0.32}_{-0.30}\times10^{9}$ $M_{sun}$ |
| $M_{*}$ | $3.7\times10^{10}$ $M_{sun}$ |
| $M_{dyn}$ | $2.7\times10^{11}$ $M_{sun}$ |
| SFR[d] | 2,900 $M_{sun}$yr$^{-1}$ |
| $\Sigma_{gas}$ | $1.4\times10^{4}$ $M_{sun}$ pc$^{-2}$ |
| $\Sigma_{SFR}$ | 600 $M_{sun}$yr$^{-1}$kpc$^{-2}$ |
| $f_{gas}$ | 40% |
| gas-to-dust ratio | 80 |
| $t_{dep}$ | 36Myr |
| $\varepsilon$ | 0.06 |
| $d_{[CII]}$ | 3.4 kpc×2.9 kpc |
| $d_{FIR}$ | 2.6 kpc×2.4 kpc |
| $T_{dust}$ | $55.9^{+9.3}_{-12.0}$K |
| $\beta$ | $1.92\pm0.12$ |

[a]assuming $\alpha_{CO} = M_{gas}/L'_{CO} = 1$ $M_{sun}$ (K kms$^{-1}$ pc$^{2}$)$^{-1}$ (refs. 27,69)
[b]assuming a [CI] excitation temperature of 30 K and a brightness temperature ratio of 0.5 between the two [CI] fine structure lines[71]
[c]based on [CII] assuming a photon dominated region (PDR) surface temperature of 144 K and a density of $10^{3.8}$ cm$^{-3}$
[d]assuming SFR[$M_{sun}$yr$^{-1}$] = $1.0\times10^{-10}$ $L_{FIR}$ [$L_{sun}$], based on a Chabrier initial mass function[62,63]

## 7. Supplementary Figures

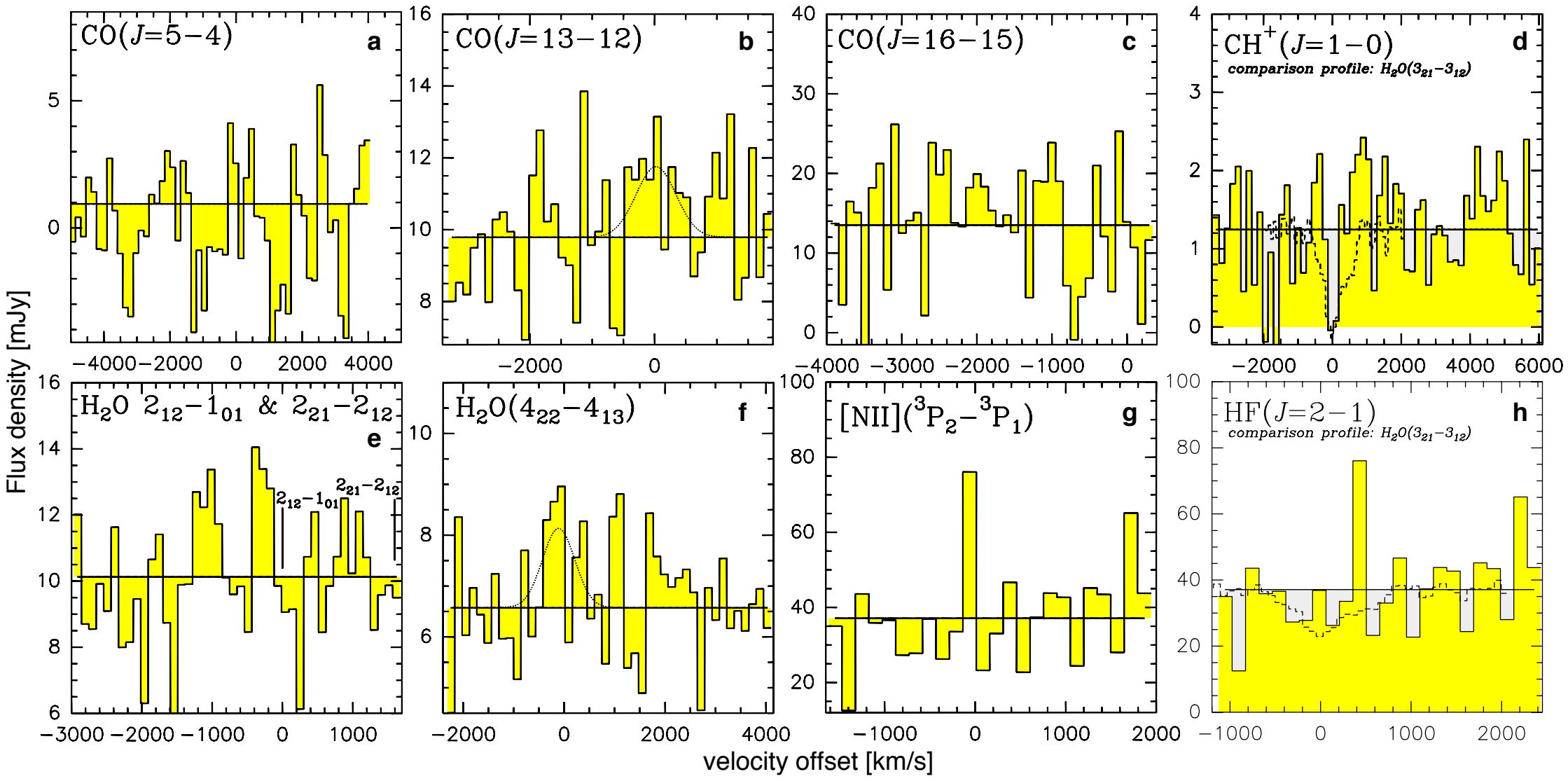

**Figure S1:** Additional diagnostic lines of the star-forming interstellar medium in HFLS3. Tentative detections and upper limits on the CO $J$=5−4, 13−12, and 16−15 lines (**a**−**c**), the $H_2O$ $J_{KaKc}$ = $2_{12}$−$1_{01}$, $2_{21}$−$2_{12}$ and $4_{22}$−$4_{13}$ lines (**e** and **f**), the $CH^+$ $J$=1−0 (**d**) and HF $J$=2−1 lines (**h**), and the [NII] $^3P_2$−$^3P_1$ line (**g**) toward HFLS3 (CO $J$=5−4 and 16−15 from CARMA, [NII] and HF from the SMA, all other tentative detections/limits from the PdBI). A more sensitive constraint on the CO $J$=16−15 line is obtained from the PdBI frequency setup targeting the OH $^2\Pi_{1/2}$ 3/2−1/2 feature. A scaled, flux-inverted profile of the $H_2O$ $J_{KaKc}$ = $3_{21}$−$3_{12}$ emission line is overplotted on the $CH^+$ $J$=1−0 and HF $J$=2−1 spectra for comparison. Although $CH^+$ is not formally detected, the spectrum shows a hint of a self-absorbed line.

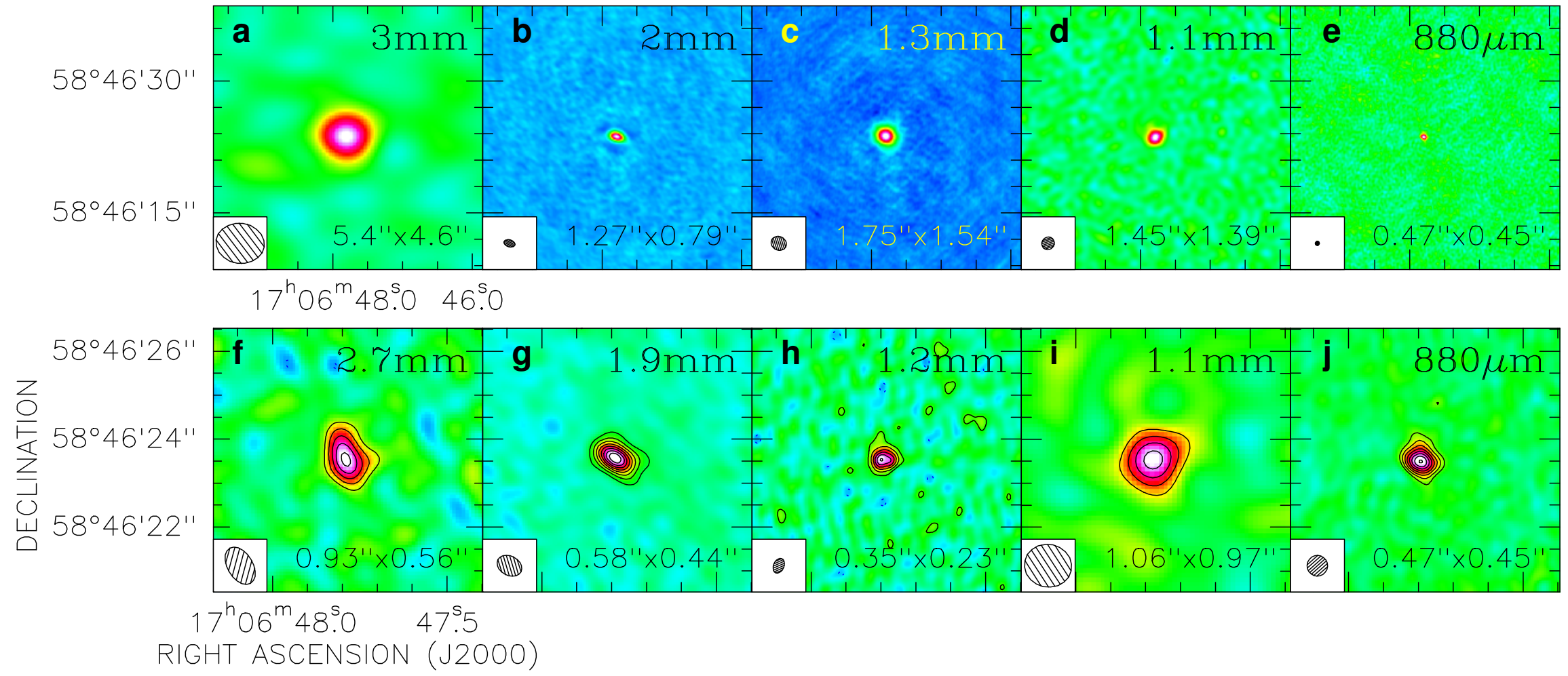

**Figure S2:** Continuum emission toward HFLS3. **a**−**e**, Maps of the observed-frame 3 mm (**a**, CARMA), 2 mm (**b**), 1.3 mm (**c**, PdBI), 1.1 mm (**d**), and 880 μm (**e**, SMA) continuum emission, as obtained by collapsing data over multiple frequency setups and array configurations. The resolution of each map is indicated in the bottom left corner of each panel. Emission is detected at 12, 148, 155, 18, and 17σ peak significance, respectively (high signal-to-noise ratio maps show beam residuals after cleaning due to limited *uv* coverage). **f-j**, High-resolution maps of the observed-frame 2.7 mm (**f**), 1.9 mm (**g**), 1.2 mm (**h**, PdBI), 1.1 mm (**i**), and 880 μm (**j**, SMA) continuum emission, as obtained by collapsing PdBI A configuration and SMA 1.1 mm EXT configuration data only (880 μm data are the same as in top row). Contours start at ±3σ (2.7 mm, 1.1 mm, and 880 μm) or ±5σ (1.9 and 1.2 mm) and are in steps of 2σ (2.7 mm, 1.1 mm, and 880 μm) or 5σ (1.9 and 1.2 mm). The emission is spatially resolved over (0.46”±0.02”)×(0.43”±0.02”), or 2.6 kpc×2.4 kpc, without evidence for emission on >~1” scales at any wavelength.

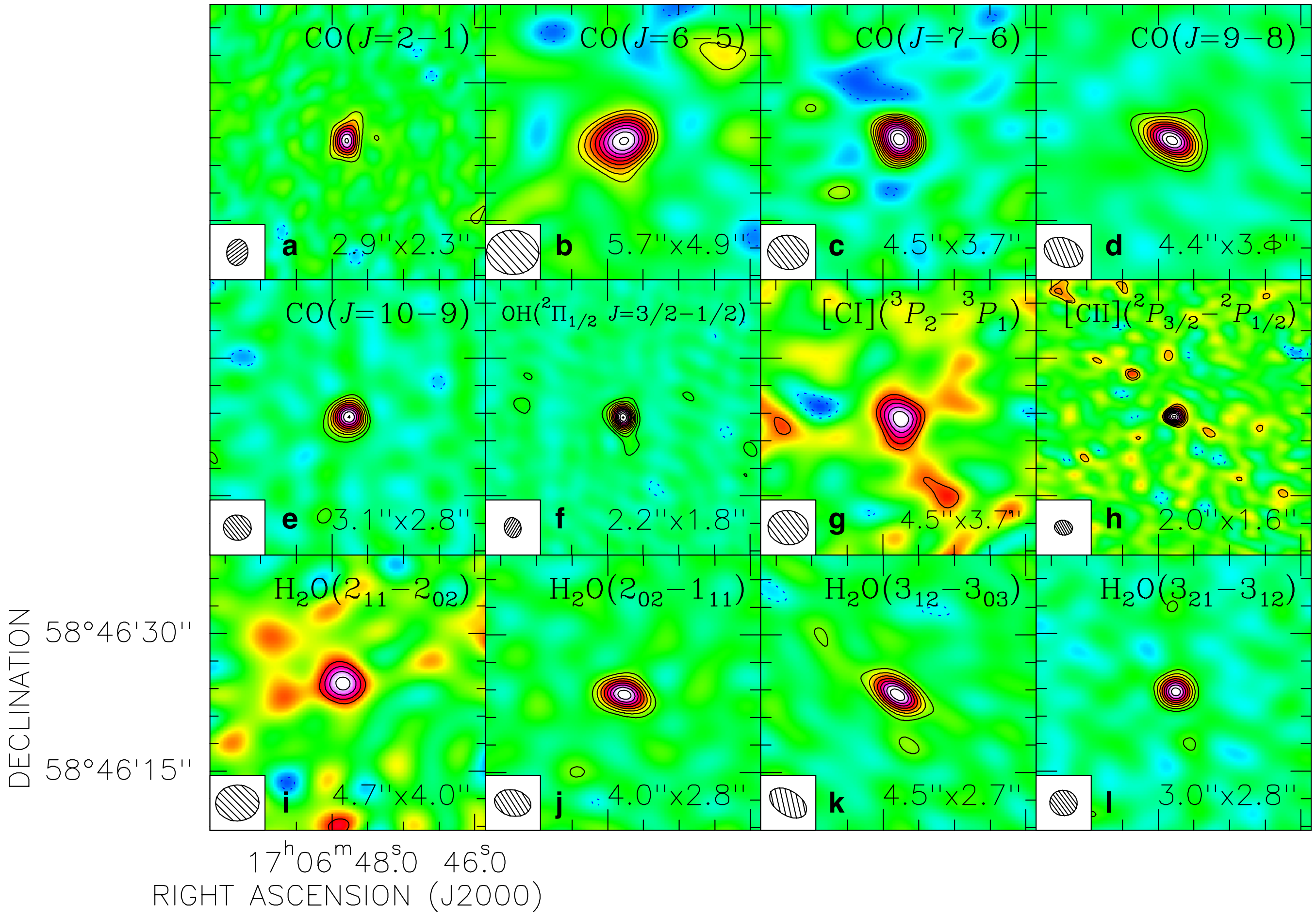

**Figure S3:** Atomic and molecular line emission towards HFLS3. CARMA, PdBI, and JVLA maps of 12 emission lines from different CO (**a**–**e**), OH (**f**), and $H_2O$ (**i**–**l**) rotational transitions and [CI] (**g**), and [CII] (**h**) fine structure lines. The beam sizes are indicated in the bottom of each panel. Contours start at ±3σ and are in steps of 1σ (CO *J*=6–5, 7–6, [CI], [CII], $H_2O$ $2_{11}$–$2_{02}$), 2σ (CO *J*=2–1, 9–8, 10–9, $H_2O$ $2_{02}$–$1_{11}$, $3_{12}$–$3_{03}$, $3_{21}$–$3_{12}$), or 5σ (OH).

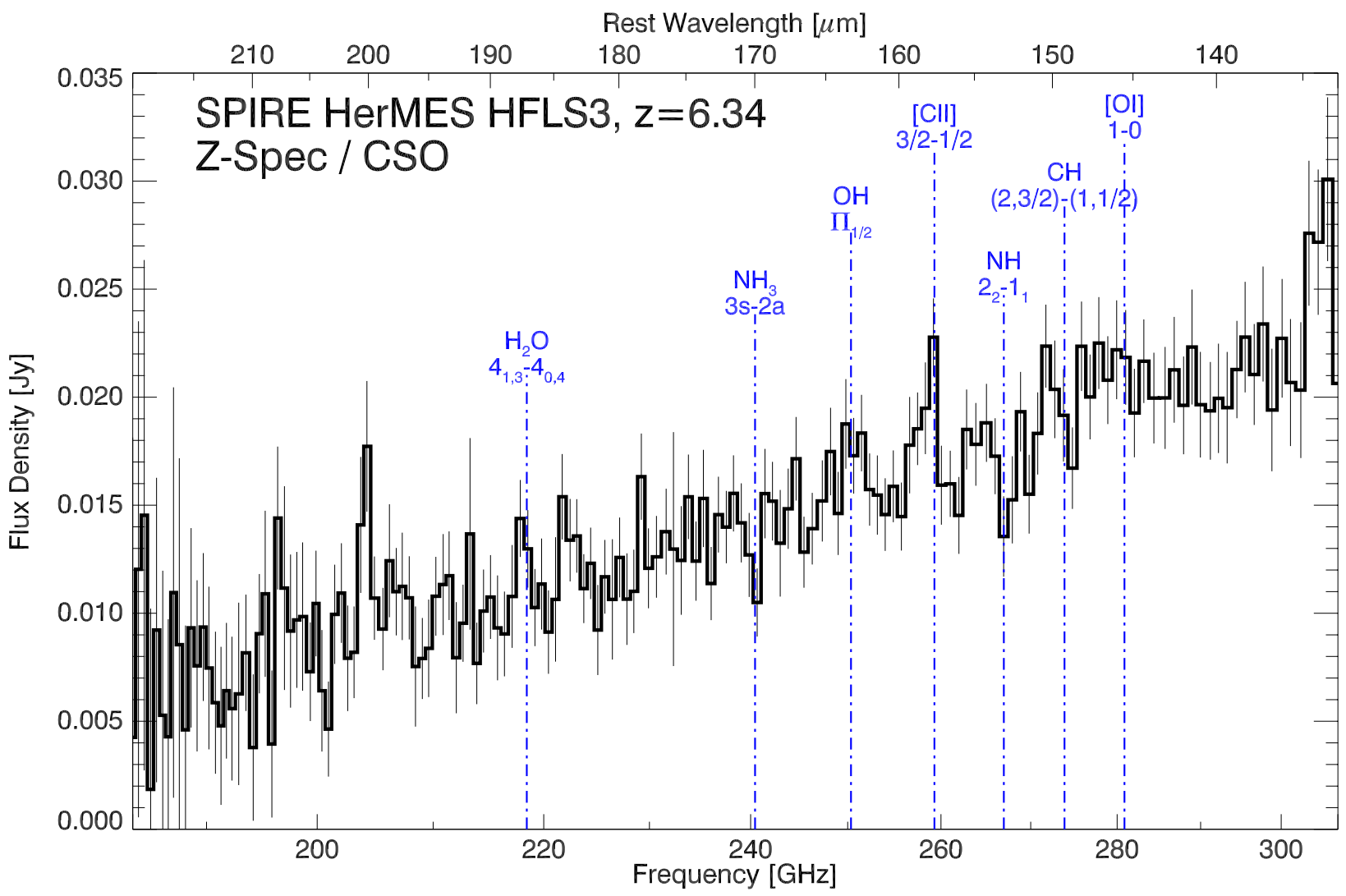

**Figure S4:** Tracers of the star-forming interstellar medium redshifted to the 1 mm window in HFLS3. CSO/Z-spec spectrum of HFLS3 with 1σ r.m.s. error bars and tentative line identifications overlayed. The [CII], OH $^2\Pi_{1/2}$ 3/2–1/2, and $NH_3$ (3,K)a–(2,K)s features were independently confirmed ($NH_3$ was only tentatively confirmed) through interferometric observations with CARMA and the PdBI. The spectrum shows an interloper line close to the redshifted frequency of CO *J*=13–12 which is not seen in interferometric observations with the PdBI (and thus, unlikely to be associated with HFLS3).

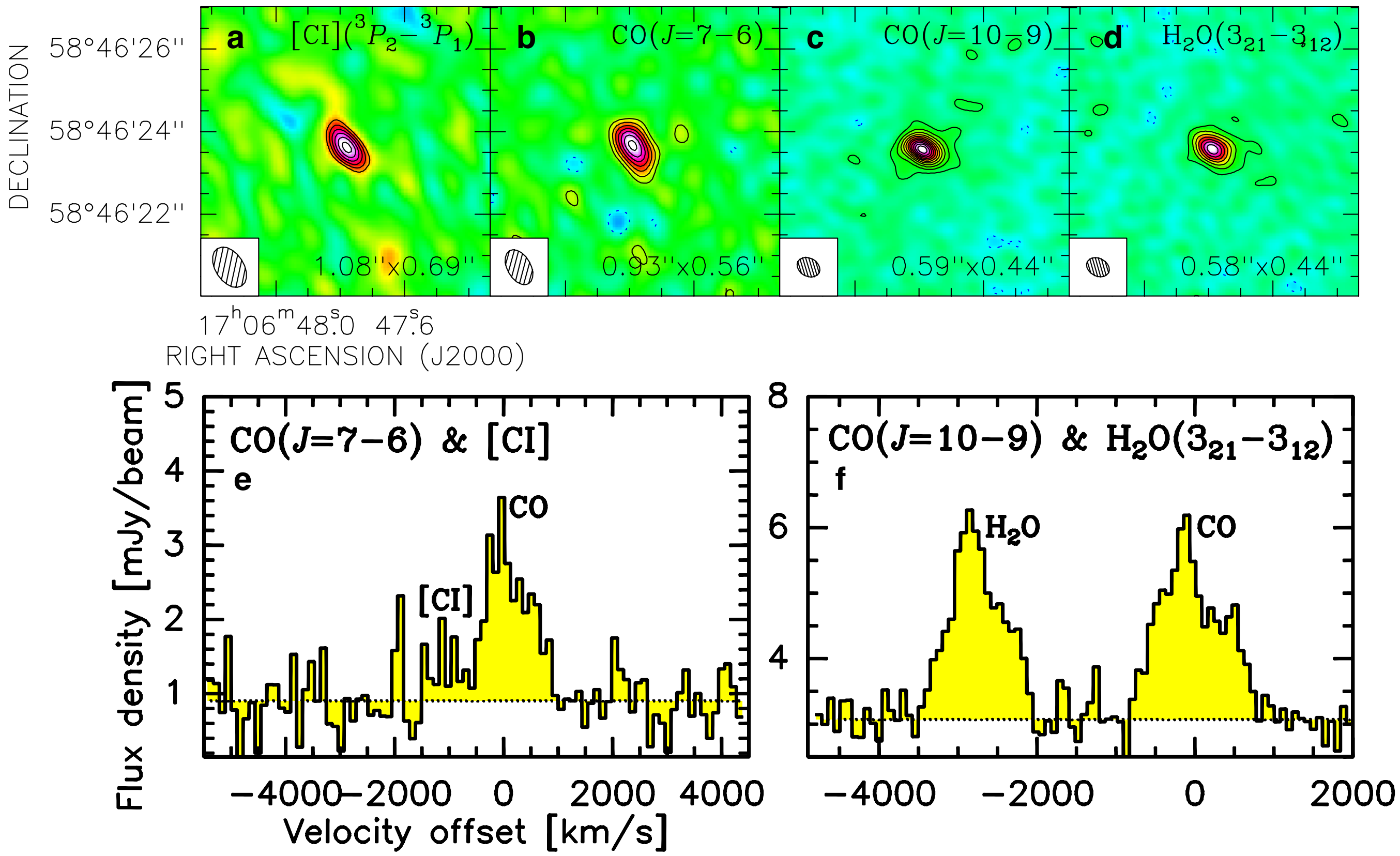

**Figure S5:** Spatially resolved atomic and molecular line emission towards HFLS3. PdBI high-resolution maps (**a**–**d**) and peak spectra (**e** and **f**) of the [CI], CO $J$=7–6 and 10–9 and $H_2O$ $3_{21}$–$3_{12}$ emission lines. The beam sizes are indicated in the bottom of each panel. Contours start at ±3σ ([CI], CO $J$=7–6 and $H_2O$) or ±5σ (CO $J$=10–9), and are in steps of 1σ ([CI]), 2σ (CO $J$=7–6), or 5σ (CO $J$=10–9 and $H_2O$).

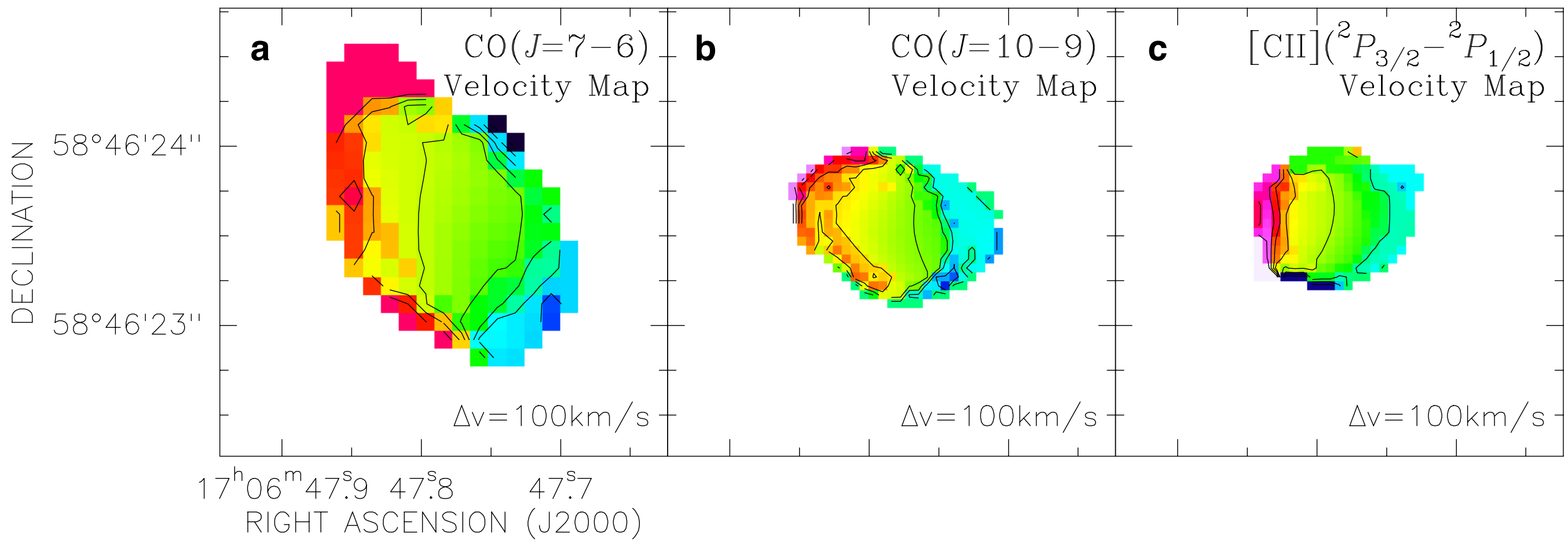

**Figure S6:** Dynamical structure of atomic and molecular line emission towards HFLS3. Velocity (first moment) maps of the CO $J$=7–6 (**a**) and 10–9 (**b**) and [CII] line emission (**c**) observed at high spatial resolution with the PdBI. Contour spacings are 100 kms$^{-1}$.

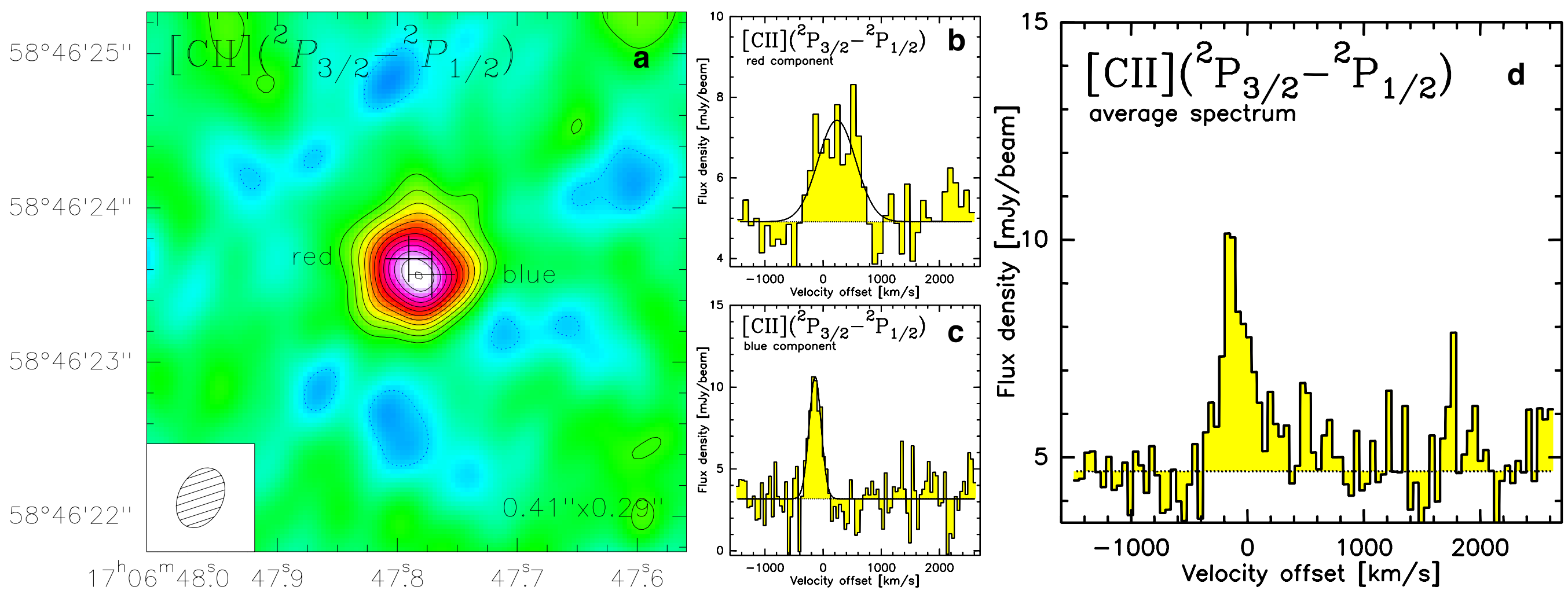

**Figure S7:** Components of [CII] line emission towards HFLS3. **a**, [CII] line map. Contour levels are in steps of 2σ, starting at ±4σ (1σ=0.33 mJy beam$^{-1}$; natural baseline weighting). **b** and **c**, [CII] velocity profiles at different positions (positions of extraction indicated by the crosses in the line map in the left panel), and average velocity profile (**d**). HFLS3 is composed of two components with velocity widths of 758±151 and 243±39 kms$^{-1}$ (FWHM) at redshifts of $z$=6.3427 and 6.3335. The varying line widths between different lines detected in this source (**Figure 1**) are due to differential excitation between these two components. The broader, "red" component peaks closer to the peak of the underlying continuum emission (**Figure 3a**), and thus, likely hosts a larger fraction of the starburst material.

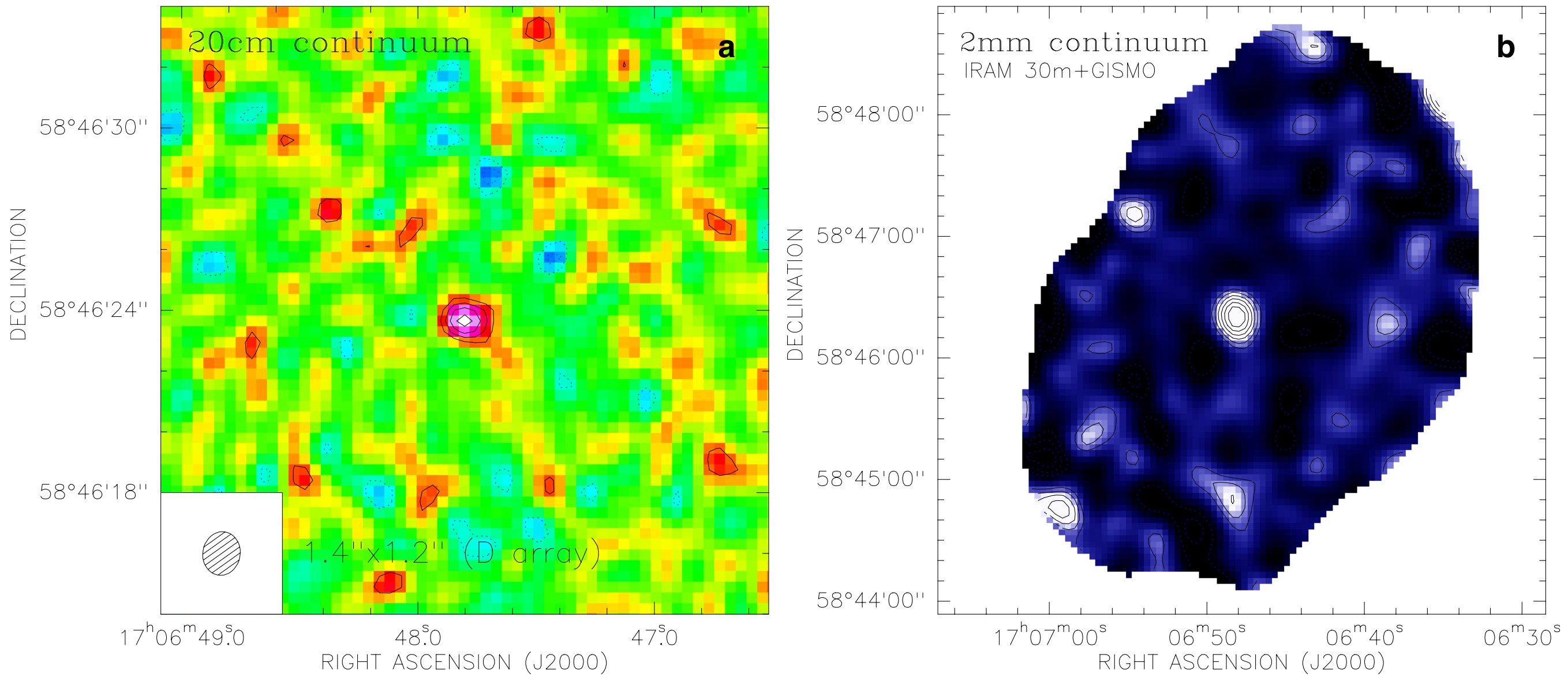

**Figure S8:** Radio continuum emission towards and region around HFLS3 at millimeter wavelengths. **a**, JVLA detection of observed-frame 20 cm (rest-frame 2.9 cm) continuum emission. Contours are in steps of 1σ = 11 μJy beam$^{-1}$, starting at ±2σ. The non-thermal synchrotron emission remains unresolved at 1.4"×1.2" (8 kpc×7 kpc) resolution, and is detected at a strength consistent with the radio-FIR correlation for nearby star-forming galaxies.[83] **b**, GISMO 2 mm continuum map of the region around HFLS3 at ~18" resolution. Contours are shown at multiples of 1σ = 0.37 mJy (negative contours are dashed).

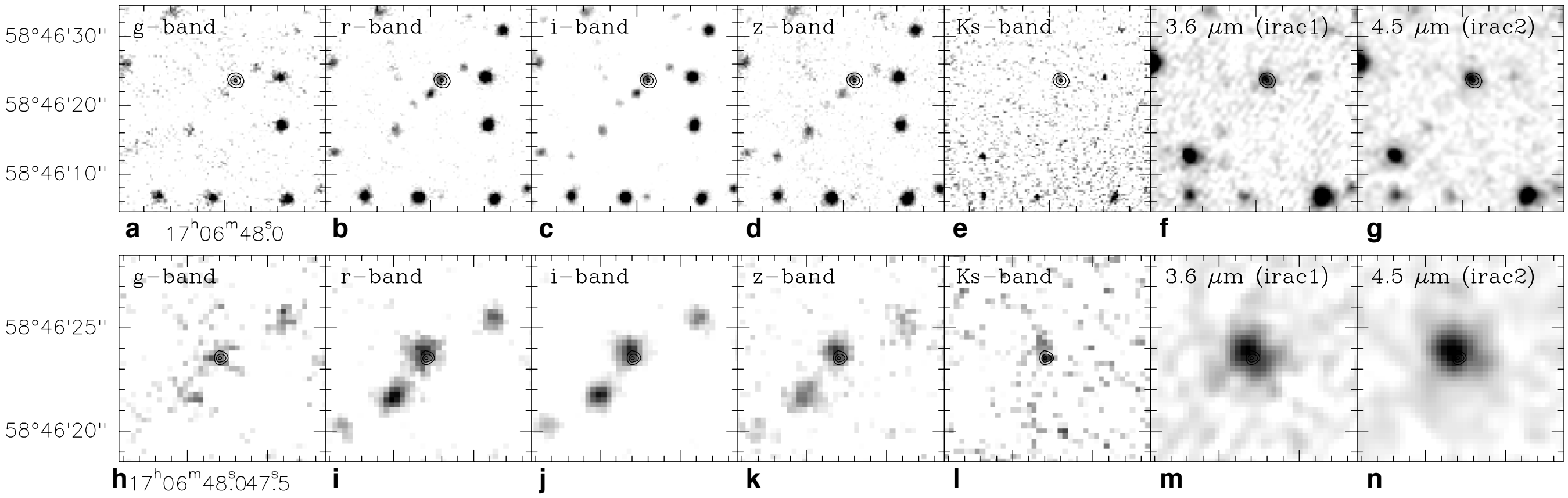

**Figure S9:** Optical to mid-infrared images of the region around HFLS3. **a–g**, 30”×30” size regions in the optical *g*, *r*, *i*, *z* (**a–d**), near-infrared $K_s$ (**e**), and mid-infrared 3.6 and 4.5 μm bands (**f** and **g**). **h–n**, zoom-in on 10”×10” size regions in the same bands. Contours of the 1 mm continuum emission are overlayed on all panels. HFLS3 is not detected in the optical bands, but is detected in $K_s$ band and longwards. The emission close to HFLS3 is dominated by the foreground galaxy G1B in all bands.

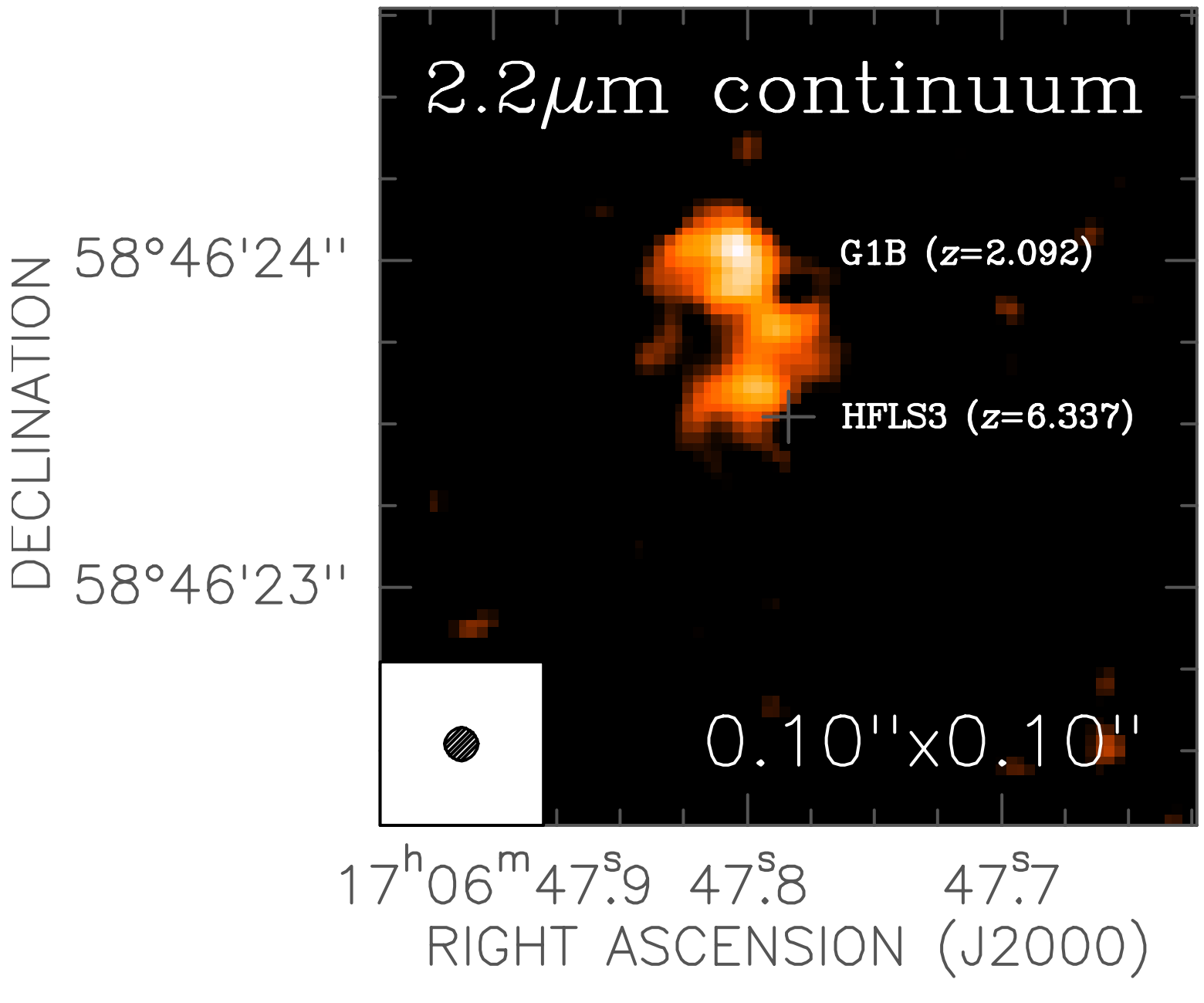

**Figure S10:** High-resolution near-infrared continuum emission towards HFLS3. Keck/NIRC2 2.2 μm Laser Guide Star Adaptive Optics image toward HFLS3, smoothed with a Gaussian with a 0.1” kernel. The image shows three bright components. The northern component is consistent with the position of the z=2.092 foreground galaxy G1B. The other two components are spatially coincident with [CII]-emitting regions in HFLS3. The south-eastern of the two components is consistent with the peak position of the observed-frame 1 mm continuum emission. Thus, the 2.2 μm observations likely detect two regions within HFLS3 with only modest dust extinction, whereas the south-western part of the [CII]-emitting region is not detected in this rest-frame 300 nm image.

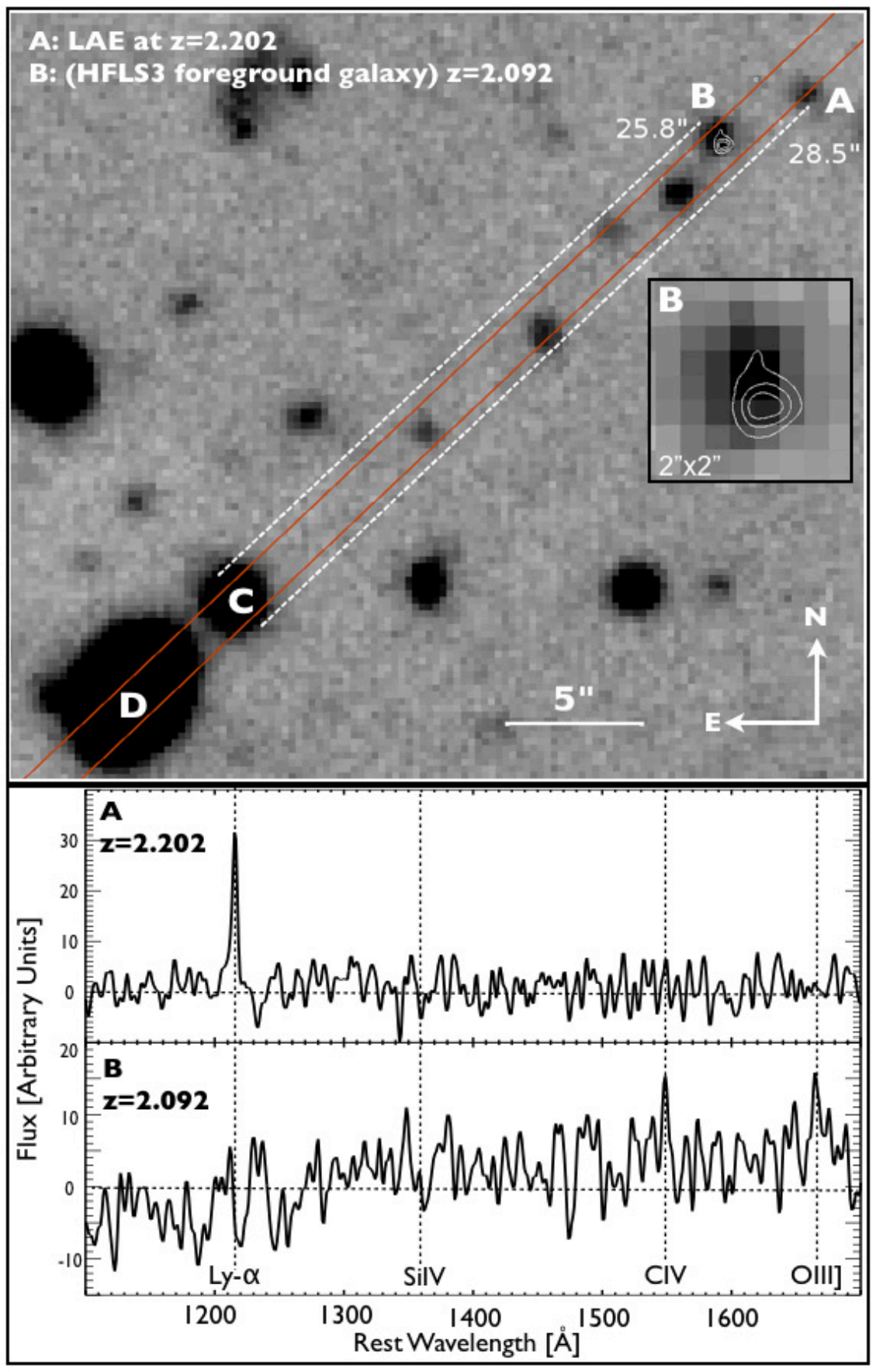

**Figure S11:** Optical spectroscopy towards HFLS3 and foreground sources. Keck/LRIS spectroscopy toward HFLS3. The top panel shows the position and alignment of the slit, overlayed on a GTC *i*-band image. The white contours correspond to the 1.2 mm continuum emission as detected toward HFLS3 with the PdBI (also shows as inset). A indicates the position of a Lyman-α emitter, B indicates the position of G1B, a galaxy ~0.65" north of HFLS3. C and D indicate reference sources. The blue lines indicate the distance of sources A and B from C. The bottom panels show the one-dimensional spectra extracted at positions A and B, which are consistent with Ly-α emission at *z*=2.202 (A) and with a Lyman-continuum break and CIV (1549 Å) and OIII] (1661,1666 Å) emission at *z*=2.092 in G1B (B). There is no indication of continuum or line emission toward HFLS3 over the entire LRIS wavelength range.

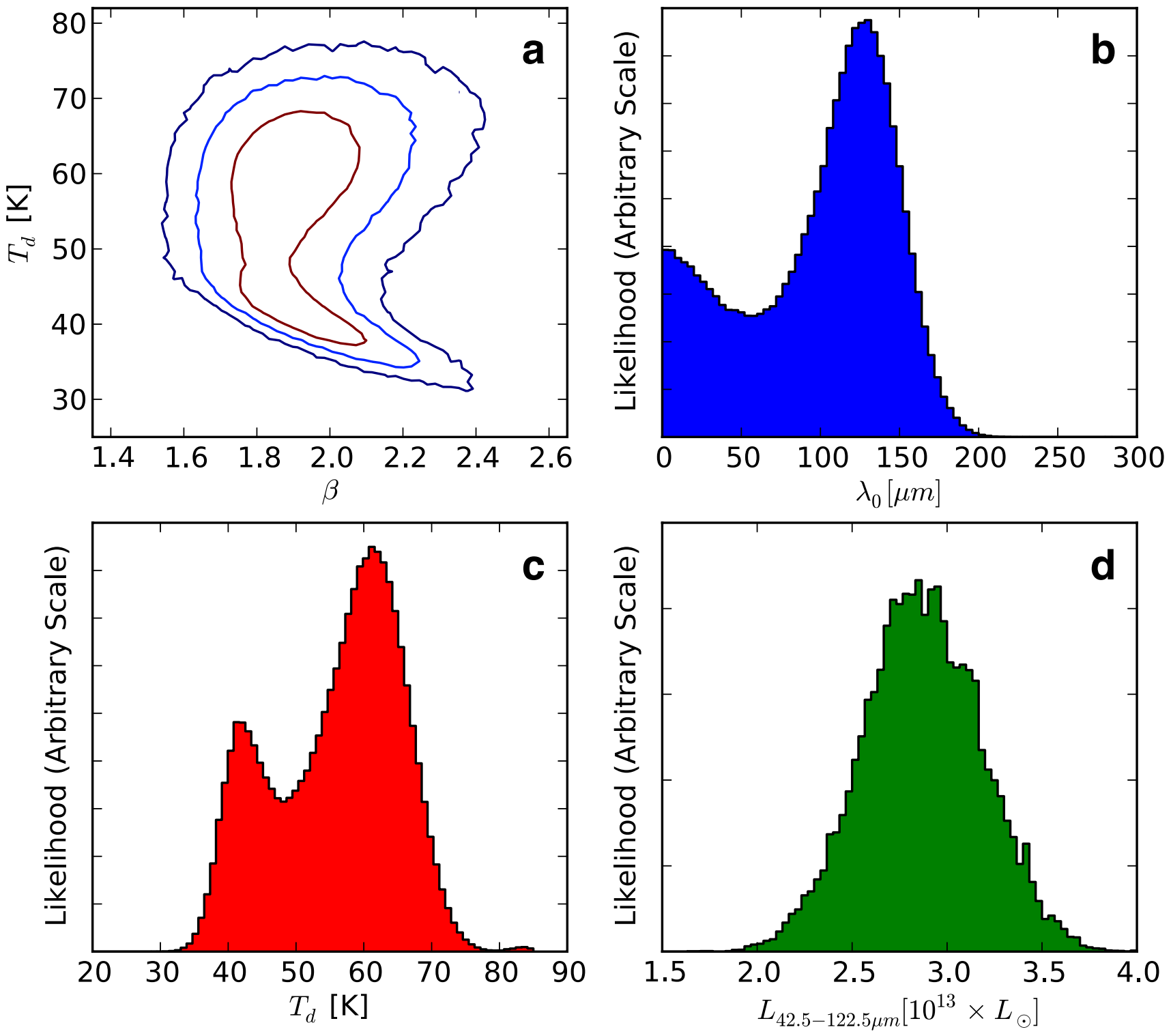

**Figure S12:** Physical parameters obtained from fitting the spectral energy distribution of HFLS3. Probabilities for $\beta$ and dust temperature (**a**, probability contours are 68.3%, 95.4%, and 99.7%), $\lambda_0$, i.e., the wavelength where the dust becomes optically thick (**b**), the dust temperature (**c**), and the far-infrared luminosity (**d**).

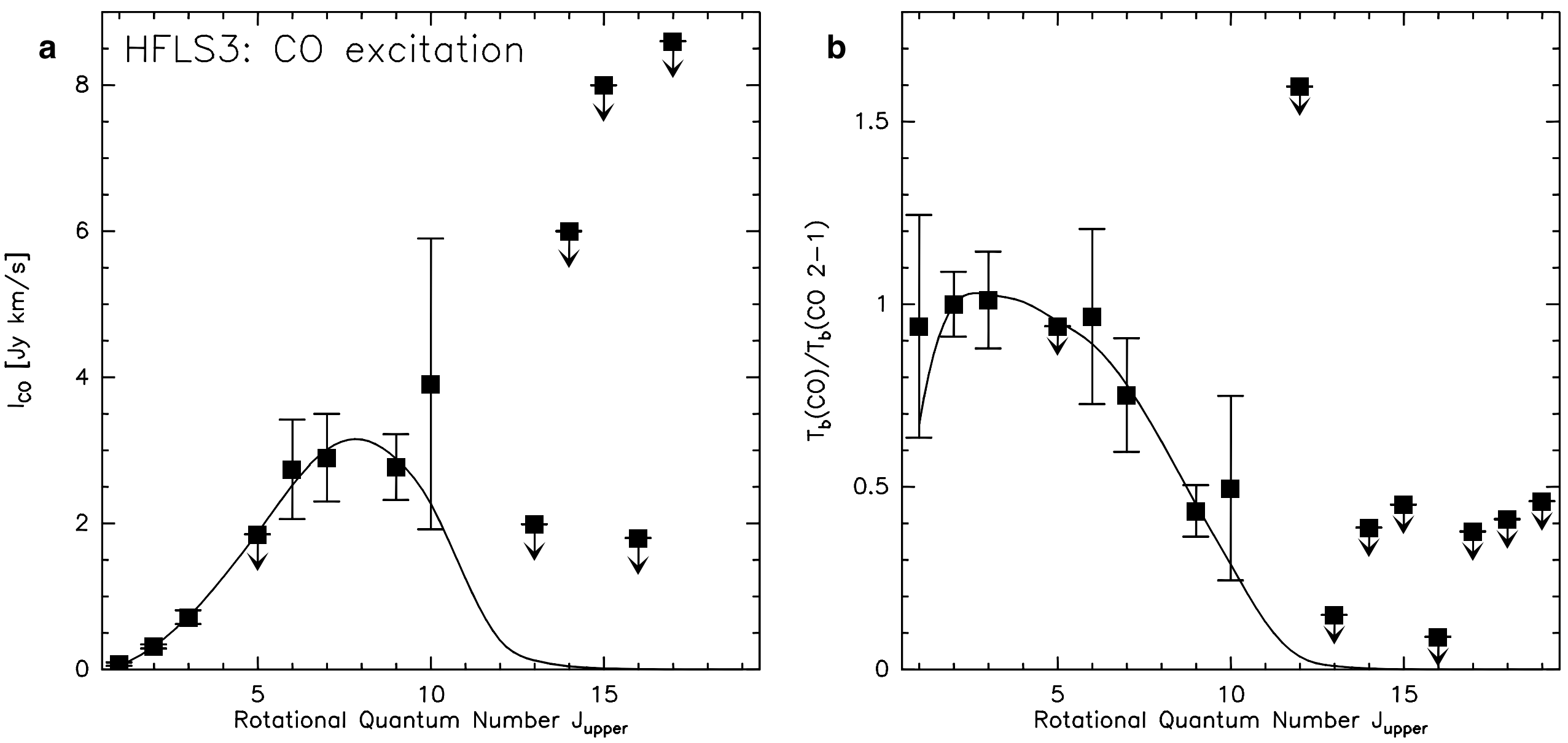

**Figure S13:** Model fits to the CO rotational line ladder of HFLS3. Observed CO excitation ladder toward HFLS3 (square symbols) and maximum likelihood RADEX model (solid line) on line intensity scale (**a**, shallow limits for some lines are off-scale), and normalized to the CO(*J*=2–1) brightness temperature (**b**).

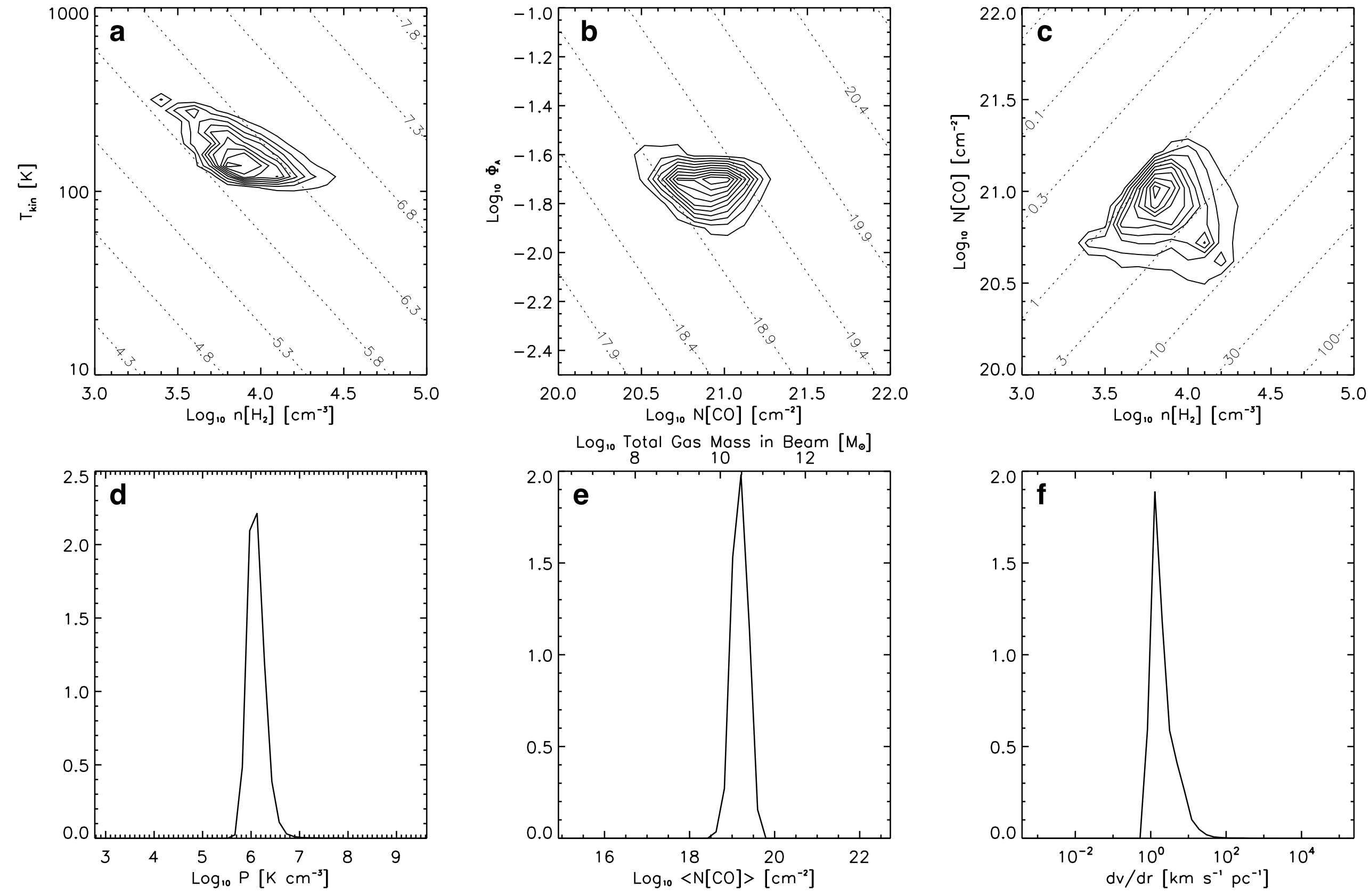

**Figure S14:** Physical parameters obtained from fitting the CO excitation ladder of HFLS3. Two-dimensional likelihood contours for the physical properties of the molecular gas in HFLS3 (**a**–**c**) as determined by RADEX models of the CO excitation ladder for $n(H_2)$ and $T_{kin}$ (**a**, dotted line shows values for *P*), $N_{CO}$ and $\Phi_A$ (**b**, dotted line shows values for <$N_{CO}$>), and $n(H_2)$ and $N_{CO}$ (**c**, dotted line shows values for d*v*/d*r*). The bottom panels show the one-dimensional likelihoods for *P* (**d**), <$N_{CO}$> (**e**), and d*v*/d*r* (**f**).

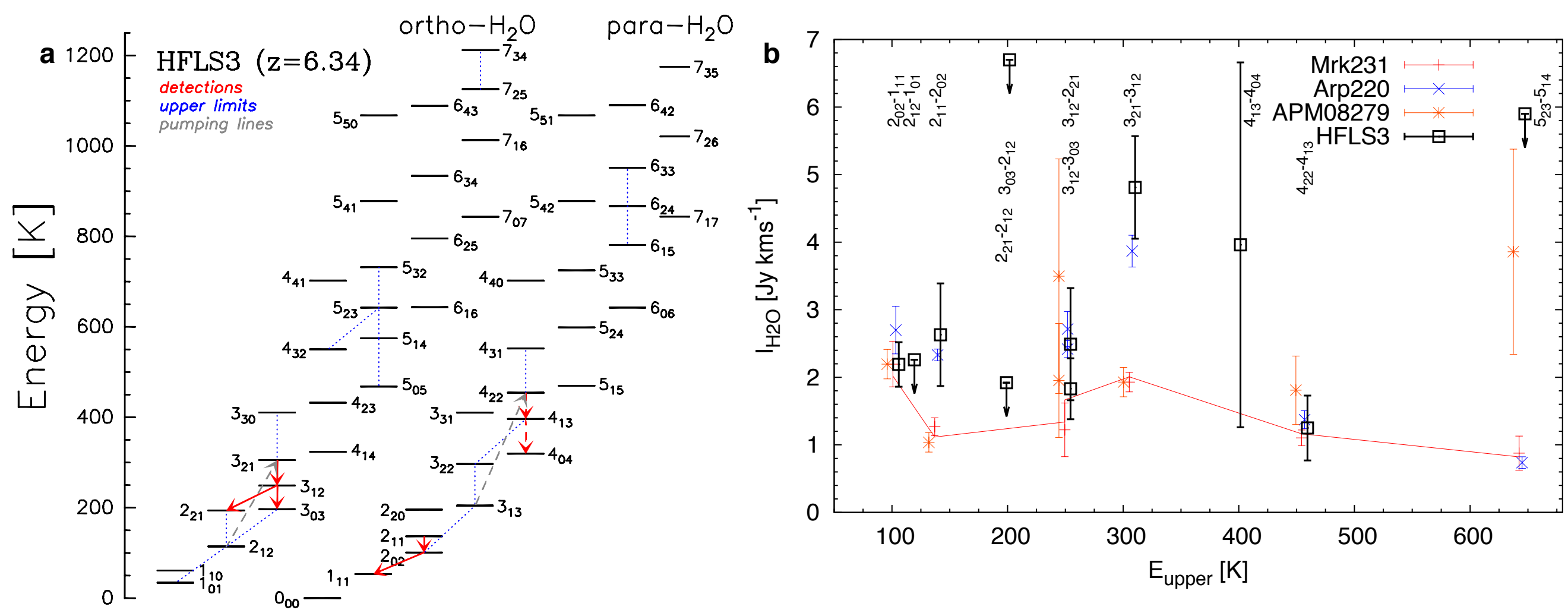

**Figure S15:** Energy level diagram (**a**) and observed line intensities (**b**) for $H_2O$ in HFLS3. **a**, The red arrows indicate detected $H_2O$ transitions (dashed lines indicate tentative detections) and blue dotted lines indicate upper limits. The dashed gray arrows indicate 75 and 58 μm infrared pumping transitions of ortho- and para-$H_2O$ that can efficiently radiatively populate the energy levels from which bright emission lines are observed. **b**, The black squares indicate observed $H_2O$ line fluxes and upper limits (shallow limits not shown) for HFLS3 as a function of upper level energy. For comparison the red, blue and orange symbols show normalized line fluxes observed in Mrk 231 (red line: model), Arp 220, and APM08279+5255.[20,21,89,90,22] Within the uncertainties, the $H_2O$ excitation in HFLS3 is comparable to that in Arp 220, but inconsistent with the line ratios in the quasars Mrk 231 and APM08279+5255 (where the $H_2O$ lines may be substantially excited by a strong X-ray radiation field associated with their AGN). In particular, the relative strength of the $J_{\mathrm{KaKc}}$=$3_{21}$–$3_{12}$ and $2_{11}$–$2_{02}$ lines relative to transitions lower in the cascade in HFLS3 suggest that their upper energy levels may be substantially populated through radiative excitation by the infrared radiation field in the star-forming regions, rather than collisions. Given the median gas density of ~$10^4$ cm$^{-3}$ as determined from the CO excitation and the critical densities of >$10^9$ cm$^{-3}$ for the detected $H_2O$ lines, contributions from collisional excitation to the observed line fluxes are likely minor. This suggests that the $H_2O$ emission lines detected in HFLS3 are less important for gas cooling through the removal of kinetic energy than the CO lines.

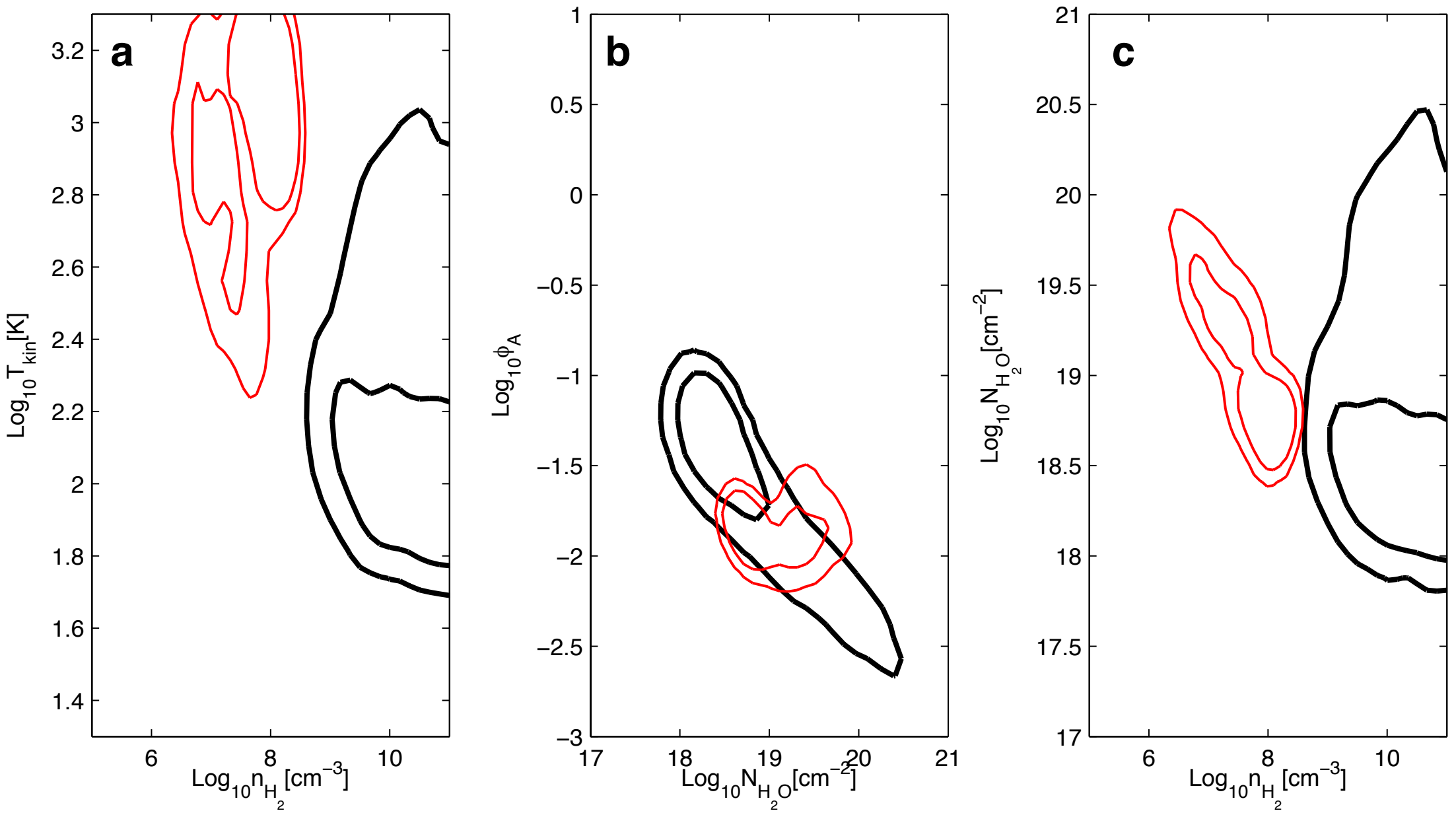

**Figure S16:** Physical parameters obtained from fitting the $H_2O$ excitation ladder of HFLS3. Two-dimensional likelihood contours for the physical properties of the molecular gas in HFLS3 as determined by RADEX models of the $H_2O$ excitation ladder for $n(H_2)$ and $T_{kin}$ (**a**), $N_{H2O}$ and $\Phi_A$ (**b**), and $n(H_2)$ and $N_{H2O}$ (**c**). Black contours show the likelihoods when fitting all observed $H_2O$ lines and limits. Red contours show the likelihoods when only considering lines with $E_{upper}/k_B<\sim200$ K. Contour levels are shown at 1 and 2σ.

## Supplementary References

31. Blain, A. *et al.* Submillimeter galaxies. *Physics Reports* **369**, 111-176 (2002).
32. Engel, H. *et al.* Most Submillimeter Galaxies are Major Mergers. *Astrophys. J.* **724**, 233-243 (2010).
33. McKee, C.F. & Ostriker, J.P. A theory of the interstellar medium - Three components regulated by supernova explosions in an inhomogeneous substrate. *Astrophys. J.* **218**, 148-169 (1977).
34. Riechers, D.A. *et al.* A Massive Molecular Gas Reservoir in the z = 5.3 Submillimeter Galaxy AzTEC-3. *Astrophys J. Lett.* **720**, L131-L136 (2010).
35. Cox, P. *et al.* Gas and Dust in a Submillimeter Galaxy at z = 4.24 from the Herschel Atlas. *Astrophys. J.* **740**, 63 (2011).
36. Combes, F. *et al.* A bright z = 5.2 lensed submillimeter galaxy in the field of Abell 773. HLSJ091828.6+514223. *Astron. Astrophys.* **538**, L4 (2012).
37. Griffin, M. *et al.* The Herschel-SPIRE instrument and its in-flight performance. *Astron. Astrophys.* **518**, L3 (2010).
38. Pilbratt, G.L. *et al.* Herschel Space Observatory. An ESA facility for far-infrared and submillimetre astronomy. *Astron. Astrophys.* **518**, L1 (2010).
39. Poglitsch, A. *et al.* The Photodetector Array Camera and Spectrometer (PACS) on the Herschel Space Observatory. *Astron. Astrophys.* **518**, L2 (2010).
40. Bradford, C.M. *et al.* Z-Spec: a broadband millimeter-wave grating spectrometer: design, construction, and first cryogenic measurements. *SPIE* **5498**, 257 (2004).
41. Bradford, C.M. *et al.* The Warm Molecular Gas around the Cloverleaf Quasar. *Astrophys. J.* **705**, 112-122 (2009).
42. Morrison, G.E. *et al.* Very Large Array 1.4 GHz Observations of the GOODS-North Field: Data Reduction and Analysis. *Astrophys. J. Suppl. Ser.* **188**, 178-186 (2010).
43. Ivison, R.J. *et al.* Tracing the molecular gas in distant submillimetre galaxies via CO(1-0) imaging with the Expanded Very Large Array. *Mon. Not. R. Astron. Soc.* **412**, 1913-1925 (2011).
44. Kovacs, A. CRUSH: fast and scalable data reduction for imaging arrays. *Proc. SPIE*, **7020**, 45 (2008).
45. Kovacs, A. SHARC-2 350 micron observations of distant submillimeter-selected galaxies and techniques for the optimal analysis and observing of weak signals. *PhD Thesis*, Caltech (2006).
46. Bertin, E. & Arnouts, S. SExtractor: Software for source extraction. *Astron. Astrophys. Suppl.* **117**, 393-404 (1996).
47. Adelman-McCarthy, J.K. *et al.* The SDSS Photometric Catalog, Release 8. *VizieR On-line Data Catalog* **II/306** (2011).
48. Cutri, R. M. *et al.* 2MASS All-Sky Catalog of Point Sources. *VizieR On-line Data Catalog* **II/246** (2003).

49. Peng, C.Y. *et al.* Detailed Structural Decomposition of Galaxy Images. *Astron. J.* **124**, 266-293 (2002).
50. Wizinowich, P.L., *et al.* The W. M. Keck Observatory Laser Guide Star Adaptive Optics System: Overview. *PASP* **118**, 297-309 (2006).
51. Oke, J.B. *et al.* The Keck Low-Resolution Imaging Spectrometer. *PASP* **107**, 375-385 (1995).
52. Wright, E.L. *et al.* The Wide-field Infrared Survey Explorer (WISE): Mission Description and Initial On-orbit Performance. *Astron. J.* **140**, 1868-1881 (2010).
53. Draine, B.T. On the Submillimeter Opacity of Protoplanetary Disks. *Astrophys. J.* **636**, 1114-1120 (2006).
54. Foreman-Mackey, D. *et al.* emcee: The MCMC Hammer. *arXiv* preprint at <http://arxiv.org/abs/1202.3665> (2012).
55. Dunne, L. *et al.* A census of metals at high and low redshift and the connection between submillimetre sources and spheroid formation. *Mon. Not. R. Astron. Soc.* **341**, 589-598 (2003).
56. Silva, L. *et al.* Modeling the Effects of Dust on Galactic Spectral Energy Distributions from the Ultraviolet to the Millimeter Band. *Astrophys. J.* **509**, 103-117 (1998).
57. Swinbank, A.M., *et al.* Intense star formation within resolved compact regions in a galaxy at z = 2.3. *Nature* **464**, 733-736 (2010).
58. Ivison, R. J., *et al.* Herschel and SCUBA-2 imaging and spectroscopy of a bright, lensed submillimetre galaxy at z = 2.3. *Astron. Astrophys.* **518**, L35 (2010).
59. Dale, D.A. & Helou, G. The Infrared Spectral Energy Distribution of Normal Star-forming Galaxies: Calibration at Far-Infrared and Submillimeter Wavelengths. *Astrophys. J.* **576**, 159-168 (2002).
60. Chary, R. & Elbaz, D. Interpreting the Cosmic Infrared Background: Constraints on the Evolution of the Dust-enshrouded Star Formation Rate. *Astrophys. J.* **556**, 562-581 (2001).
61. Conley, A. *et al.* Discovery of a Multiply Lensed Submillimeter Galaxy in Early HerMES Herschel/SPIRE Data. *Astrophys. J. Lett.* **732**, L35 (2011).
62. Chabrier, G. The Galactic Disk Mass Function: Reconciliation of the Hubble Space Telescope and Nearby Determinations. *Astrophys. J.* **586**, L133-L136 (2003).
63. Kennicutt, R.C. Jr. The Global Schmidt Law in Star-forming Galaxies. *Astrophys. J.* **498**, 541-552 (1998).
64. Baugh, C.M. *et al.* Can the faint submillimetre galaxies be explained in the Λ cold dark matter model? *Mon. Not. R. Astron. Soc.* **356**, 1191-1200 (2005).
65. Swinbank, A.M. *et al.* The properties of submm galaxies in hierarchical models. *Mon. Not. R. Astron. Soc.* **391**, 420-434 (2008).
66. Cappellari, M. *et al.* Systematic variation of the stellar initial mass function in early-type galaxies. *Nature*, **484**, 485-488 (2012).
67. van Dokkum, P.G. & Conroy, C. A substantial population of low-mass stars in luminous elliptical galaxies. *Nature*, **468**, 940-942 (2010).
68. Conroy, C. & van Dokkum, P.G. The Stellar Initial Mass Function in Early-type Galaxies From Absorption Line Spectroscopy. II. Results. *Astrophys. J.* **760**, 71 (2012).
69. Tacconi, L.J., *et al.* Submillimeter Galaxies at z ~ 2: Evidence for Major Mergers and Constraints on Lifetimes, IMF, and CO-$H_2$ Conversion Factor. *Astrophys. J.* **680**, 246-262 (2008).
70. Daddi, E. *et al.* Very High Gas Fractions and Extended Gas Reservoirs in z = 1.5 Disk Galaxies. *Astrophys. J.* **713**, 686-707 (2010).
71. Walter, F. *et al.* A Survey of Atomic Carbon at High Redshift. *Astrophys. J.* **730**, 18 (2011).
72. Hailey-Dunsheath, S. *et al.* Detection of the 158 μm [C II] Transition at z = 1.3: Evidence for a Galaxy-wide Starburst. *Astrophys. J. Lett.* **714**, L162-L166 (2010).
73. Bruzual , G. & Charlot S. Stellar population synthesis at the resolution of 2003. *Mon. Not. R. Astron. Soc.* **344**, 1000-1028 (2003).
74. Bolzonella, M. *et al.* Photometric redshifts based on standard SED fitting procedures. *Astron. Astrophys.* **363**, 476-492 (2000).
75. Giovannoli, E. *et al.* Population synthesis modelling of luminous infrared galaxies at intermediate redshift. *Astron. Astrophys.* **525**, A150 (2011).
76. Pforr, J. *et al.* Recovering galaxy stellar population properties from broad-band spectral energy distribution fitting. *Mon. Not. R. Astron. Soc.* **422**, 3285-3326 (2012).
77. de Barros, S. *et al.* Properties of z ~ 3 to z ~ 6 Lyman Break Galaxies. I. Impact of nebular emission at high redshift. *arXiv* preprint at <http://arxiv.org/abs/1207.3663> (2012).
78. Hainline, L.J. *et al.* The Stellar Mass Content of Submillimeter-selected Galaxies. *Astrophys. J.* **740**, 96 (2011).
79. Bothwell, M.S. *et al.* A survey of molecular gas in luminous sub-millimetre galaxies. Submitted to *Mon. Not. R. Astron. Soc.; arXiv* preprint at <http://arxiv.org/abs/1205.1511> (2012).
80. Condon, J.J. Radio emission from normal galaxies. *Ann. Rev. Astron. Astrophys.* **30**, 575-611 (1992).
81. Ibar, E. *et al.* Deep multi-frequency radio imaging in the Lockman Hole - II. The spectral index of submillimetre galaxies. *Mon. Not. R. Astron. Soc.* **401**, L53-L57 (2010).
82. Helou, G. *et al.* Thermal infrared and nonthermal radio - Remarkable correlation in disks of galaxies. *Astrophys. J. Lett.* **298**, L7-L11 (1985).
83. Yun, M.S., *et al.* Radio Properties of Infrared-selected Galaxies in the IRAS 2 Jy Sample. *Astrophys. J.* **554**, 803-822 (2001).

84. Ivison, R.J. *et al.* The far-infrared/radio correlation as probed by Herschel. *Astron. Astrophys.* **518**, L31 (2010).
85. van der Tak, F.F.S. *et al.* A computer program for fast non-LTE analysis of interstellar line spectra. With diagnostic plots to interpret observed line intensity ratios. *Astron. Astrophys.* **468**, 627-635 (2007).
86. Kamenetzky, J. *et al.* The Dense Molecular Gas in the Circumnuclear Disk of NGC 1068. *Astrophys. J.* **731**, 83 (2011).
87. Feroz, F. & Hobson, M.P. Multimodal nested sampling: an efficient and robust alternative to Markov Chain Monte Carlo methods for astronomical data analyses. *Mon. Not. R. Astron. Soc.* **384**, 449-463 (2008).
88. Harris, A.I. *et al.* Blind detections of CO J = 1-0 in 11 H-ATLAS galaxies at z = 2.1-3.5 with the GBT/Zpectrometer. *Astrophys. J.* **752**, 152 (2012).
89. van der Werf, P.P. *et al.* Black hole accretion and star formation as drivers of gas excitation and chemistry in Markarian 231. *Astron. Astrophys.* **518**, L42 (2010).
90. Bradford, C.M. *et al.* The Water Vapor Spectrum of APM 08279+5255: X-Ray Heating and Infrared Pumping over Hundreds of Parsecs. *Astrophys. J. Lett.* **741**, L37 (2011).

## Acknowledgments

We thank Lee Armus and Tanio Diaz-Santos for help with determining the stellar mass of Arp 220. SPIRE has been developed by a consortium of institutes led by Cardiff Univ. (UK) and including Univ. Lethbridge (Canada); NAOC (China); CEA, LAM (France); IFSI, Univ. Padua (Italy); IAC (Spain); Stockholm Observatory (Sweden); Imperial College London, RAL, UCL-MSSL, UKATC, Univ. Sussex (UK); Caltech, JPL, NHSC, Univ. Colorado (USA). This development has been supported by national funding agencies: CSA (Canada); NAOC (China); CEA, CNES, CNRS (France); ASI (Italy); MCINN (Spain); SNSB (Sweden); STFC (UK); and NASA (USA). PACS has been developed by a consortium of institutes led by MPE (Germany) and including UVIE (Austria); KU Leuven, CSL, IMEC (Belgium); CEA, LAM (France); MPIA (Germany); INAF-IFSI/OAA/OAP/OAT, LENS, SISSA (Italy); IAC (Spain). This development has been supported by the funding agencies BMVIT (Austria), ESA-PRODEX (Belgium), CEA/CNES (France), DLR (Germany), ASI/INAF (Italy), and CICYT/MCYT (Spain). Support for CARMA construction was derived from the states of California, Illinois, and Maryland, the James S. McDonnell Foundation, the Gordon and Betty Moore Foundation, the Kenneth T. and Eileen L. Norris Foundation, the University of Chicago, the Associates of the California Institute of Technology, and the National Science Foundation. Ongoing CARMA development and operations are supported by the National Science Foundation under a cooperative agreement, and by the CARMA partner universities. The CSO is funded by the NSF under contract AST 02-29008. Based on observations carried out with the IRAM Plateau de Bure Interferometer. IRAM is supported by INSU/CNRS (France), MPG (Germany) and IGN (Spain). The National Radio Astronomy Observatory is a facility of the National Science Foundation operated under cooperative agreement by Associated Universities, Inc. The Submillimeter Array is a joint project between the Smithsonian Astrophysical Observatory and the Academia Sinica Institute of Astronomy and Astrophysics and is funded by the Smithsonian Institution and the Academia Sinica. The GISMO observations were partially funded through NSF ATI grants 1020981 and 1106284. Some of the data presented herein were obtained at the W.M. Keck Observatory, which is operated as a scientific partnership among the California Institute of Technology, the University of California and the National Aeronautics and Space Administration. The Observatory was made possible by the generous financial support of the W.M. Keck Foundation. The authors wish to recognize and acknowledge the very significant cultural role and reverence that the summit of Mauna Kea has always had within the indigenous Hawaiian community. We are most fortunate to have the opportunity to conduct observations from this mountain. Based on observations made with the William Herschel Telescope (WHT), in part under Director's Discretionary Time of Spain's Instituto de Astrofisica de Canarias (IAC), and with the Gran Telescopio Canarias (GTC), installed in the Spanish Observatorio del Roque de los Muchachos of the IAC, in the island of La Palma. The WHT is operated on the island of La Palma by the Isaac Newton Group. The GTC and some of the WHT observations are part of the International Time Programme 2010–2011 (PI: Perez-Fournon). This publication makes use of data products from the Wide-field Infrared Survey Explorer, which is a joint project of the University of California, Los Angeles, and the Jet Propulsion Laboratory/California Institute of Technology, funded by the National Aeronautics and Space Administration.